\documentclass[journal]{IEEEtran}
\usepackage{cite}
\usepackage{url}

\usepackage{textcomp}
\usepackage{xcolor}
\usepackage{hyperref}

\usepackage{booktabs}	
\usepackage{diagbox}    
\usepackage{multirow}   

\usepackage{verbatim}	

\usepackage{amsmath,amssymb,amsfonts,graphicx}
\usepackage{epstopdf}

\usepackage{algorithm}
\usepackage{algorithmicx}
\usepackage{algpseudocode}

\newtheorem{definition}{Definition}[section]

\newtheorem{strategy}{Strategy}    

\renewcommand{\algorithmicrequire}{\textbf{Input:}} 
\renewcommand{\algorithmicensure}{\textbf{Output:}} 
\def\BibTeX{{\rm B\kern-.05em{\sc i\kern-.025em b}\kern-.08em
		T\kern-.1667em\lower.7ex\hbox{E}\kern-.125emX}}

\usepackage{amssymb}
\usepackage{hyperref}

\usepackage[switch]{lineno}   

\ifCLASSINFOpdf
\else
\fi

\begin{document}

\title{THUE: Discovering Top-$K$ High Utility Episodes}

\author{Shicheng Wan, Jiahui Chen,~\IEEEmembership{Member,~IEEE,} 
	 Wensheng Gan,~\IEEEmembership{Member,~IEEE,}  \\    
	Guoting Chen, and Vikram Goyal
	

\thanks{This work was partially supported by the National Natural Science Foundation of China under Grants 61902079 and 62002136, the Key Areas Research and Development Program of Guangdong Province under Grant 2019B010139002, and the project of Guangzhou Science and Technology under Grants 201902020006 and 201902020007. (Corresponding author: Wensheng Gan)}
	
	\thanks{Shicheng Wan and Jiahui Chen  are with the Department of Computer Sciences, Guangdong University of Technology, Guangzhou 510006, China. (E-mail: scwan1998@gmail.com and csjhchen@gmail.com)}

	\thanks{Wensheng Gan is with the College of Cyber Security, Jinan University, Guangzhou 510632, Guangdong, China; and with Guangdong Artificial Intelligence and Digital Economy Laboratory (Pazhou Lab), Guangzhou 510335, China. (E-mail: wsgan001@gmail.com)}

	\thanks{Guoting Chen is with the School of Science, Harbin Institute of Technology (Shenzhen), Shenzhen, China (E-mail: chenguoting@hit.edu.cn)}

	\thanks{Vikram Goyal is with Department of Computer Science, IIT-Delhi, New Delhi, India (E-mail: vikram@iiitd.ac.in)}

}

\maketitle


\begin{abstract}
	Episode discovery from an event is a popular framework for data mining tasks and has many real-world applications. An episode is a partially ordered set of objects (e.g., item, node), and each object is associated with an event type. This episode can also be considered as a complex event sub-sequence. High-utility episode mining is an interesting utility-driven mining task in the real world. Traditional episode mining algorithms, by setting a threshold, usually return a huge episode that is neither intuitive nor saves time. In general, finding a suitable threshold in a pattern-mining algorithm is a trivial and time-consuming task. In this paper, we propose a novel algorithm, called \textbf{T}op-$K$ \textbf{H}igh \textbf{U}tility \textbf{E}pisode (THUE) mining within the complex event sequence, which redefines the previous mining task by obtaining the $k$ highest episodes. We introduce several threshold-raising strategies and optimize the episode-weighted utilization upper bounds to speed up the mining process and effectively reduce the memory cost. Finally, the experimental results on both real-life and synthetic datasets reveal that the THUE algorithm can offer six to eight orders of magnitude running time performance improvement over the state-of-the-art algorithm and has low memory consumption.
\end{abstract}


\IEEEpeerreviewmaketitle

\section{Introduction}

\IEEEPARstart{I}{n} the past decades, a multitude of investigators have hastened to improve frequent pattern mining (FPM) in the data mining field \cite{mannila1995discovering,mannila1997discovery,gan2017data,geng2006interestingness}. As a significant part of the domain of FPM, frequent episode mining (FEM) has been studied \cite{mannila1995discovering,mannila1997discovery,ma2004finding,huang2008efficient}. An episode is a set of partially ordered events that are described by their types, and each event is always associated with an occurrence time point \cite{mannila1997discovery,laxman2007discovering}. In essence, an episode is a sequence of symbols, and we also assume that they are a fundamental type of data. It aims to determine the relationships between some events, which can be divided into three categories: 1) serial episodes, denoted as a sequence of event types; 2) parallel episodes, denoted as a set of event types; and 3) complex episodes, denoted as a mixture of serial and parallel episodes. Applications of FEM include stock trend prediction \cite{ng2003mining,dattasharma2008identifying}, customer behavior analysis \cite{mannila1997discovery}, and prediction of user behavior in web click streams \cite{laxman2007fast}.

Although FEM has been successfully applied in many applications, it has an inevitable shortcoming. The frequency of a pattern cannot completely reveal the interest of a pattern; it only considers the occurrence and absence of events, but ignores the information of different types of events \cite{gan2020fast}. Furthermore, most studies on FEM can only solve simple event sequences, and few consider complex event sequences where different types of events occur simultaneously at the same timestamp. Thus, researchers have focused on discovering useful knowledge with utility pattern mining (UPM) \cite{geng2006interestingness,liu2005two,gan2018survey,gan2021survey} framework. In UPM, utility is a new measurement, replacing the frequency metric. It consists of external utility (e.g., unit profit) and internal utility (e.g., purchase quantity). If the utility of a pattern in a dataset is no less than the user-specified minimum utility (\textit{minUtil}) threshold, then we suppose that it is a high-utility pattern. In practice, there are many applications in UPM with episodes, such as website click stream analysis \cite{ahmed2011framework}, stock investment \cite{lin2017novel}, cross-marketing in retail stores \cite{guo2014high,rathore2016top}, and cloud workload prediction \cite{amiri2018online}. Taking stock investment as a sample, consider the following financial report \cite{ng2003mining}: ``Telecommunications stocks pushed the Hang Seng Index 2\% higher following the Star TV--HK Telecom and Orange--Mannesmann deals." ``Star TV--HK Telecom + Orange--Mannesmann $\Rightarrow$ telecommunications stocks pushed + Hang Seng Index 2\% higher" can be viewed as a complex episode. In the FEM field, it only counts the occurrence times of this episode and then predicts the stock's trend. However, UPM algorithms regard these events (``telecommunication stocks rise," ``Hang Seng Index surges," ``Star TV--HK Telecom,'' and ``Orange--Mannesmann") with distinct weights/utilities. The ``Star TV--HK Telecom'' event may play a more important role than ``Orange--Mannesmann" in the episode, and may be the key reason that ``telecommunication stocks rise" occurred. However, the frequency metric cannot achieve this goal.

FEM generally employs the monotone/anti-monotone property (\textit{downward closure property}) to effectively prune a large search space. However, the \textit{downward closure property} does not hold in UPM directly, which indicates that the previous optimization methods in FEM become invalid. Fortunately, \textit{ transaction weighted utilization } (\textit{TWU}) \cite{liu2005two} has been previously proposed to successfully tackle this problem. With the \textit{TWU} model, if the \textit{TWU} of a pattern is no less than the user-defined \textit{minUtil} threshold, then we suppose that it is a potential high-utility pattern, and check its real utility to determine whether it is still greater than the \textit{minUtil} threshold in the next step. The problem of finding interesting episodes according to their utility is formulated as high-utility episode mining (HUEM) \cite{guo2012a}. The next challenge is to incorporate HUEM with the \textit{TWU} model. Since the episode database can be considered as a single and very long event sequence, it is completely different from the transaction database and sequence database. Wu \textit{et al.} \cite{wu2013mining} proposed a novel concept called \textit{ episode weighted utilization } (\textit{EWU}), which is similar to \textit{TWU} and plays the same role as complex event sequences. They discovered high-utility episodes (HUEs) that meet the maximum time duration (\textit{MTD}) constraint, while their utility values are greater than the \textit{minUtil} threshold. Subsequently, Guo \textit{et al.} \cite{guo2014high} proposed the TSpan algorithm, which utilizes a lexicographic sequence tree to find HUEs efficiently. However, both works require a large amount of memory and runtime consumption while mining. Fortunately, Gan \textit{et al.} \cite{gan2019discovering} not only obtained accurate interesting episodes, but also achieved a desirable operational efficiency. Nevertheless, all the aforementioned algorithms meet the same limitation that the \textit{TWU} model faced. In the HUEM domain, the quantity of HUEs finally outputs, which is deeply influenced by the value of the \textit{minUtil} threshold and the characteristics of the database. Setting an appropriate \textit{minUtil} threshold is a key task. It is quite possible that we obtain few episodes if it is set high, whereas if the threshold is too low, many uninteresting episodes result in redundant information. To set a suitable threshold, users must perform a detailed analysis of the items, utility values, and characteristics of the databases. Obviously, this is a disturbing task. Furthermore, this phenomenon can be described as a question: How many HUEs do we really need? In general, managers are more likely to learn about the top-$k$ profitable goods instead of hundreds of thousands of results in retail stores \cite{zhang2020tkus}. Fortunately, Rathore \textit{et al.} \cite{rathore2016top} completed this hard work in recent years. The parameter $k$ is the number of HUEs that users need. The \textit{minUtil} threshold usually starts at 1 and increases automatically during the discovery of HUEs.

In this paper, we propose a new efficient algorithm, named THUE, for mining top-$k$ HUEs from a complex event sequence. It is updated using the UMEpi algorithm \cite{gan2019discovering}. The experimental results show that THUE not only obtains the top-$k$ correct HUEs, but also does not miss any HUEs. Our main contributions can be summarized as follows.

\begin{itemize}
	\item We utilize a new definition of EWU, which can be more accurate than the existing top-$k$ HUEM method to prune the search space.
	
	\item There is no need to consider different applicable situations separately. The proposed algorithm has good flexibility to process different types of event sequences, such as simultaneous, serial, or complex events.
	
	\item We develop several effective strategies to raise the \textit{minUtil} threshold as soon as possible during the mining process.
	
	\item We test real and synthetic databases and compare the performance with the state-of-the-art TUP \cite{rathore2016top} algorithm. The proposed algorithm generates fewer candidates and is at least three times faster than TUP.
	
\end{itemize}

The remainder of this paper is organized as follows. Related works are briefly reviewed in Section \ref{sec:related}, and Section \ref{sec:preliminaries} introduces the key preliminaries and main problem statements of HUEM. Section \ref{sec:algorithms} describes the proposed THUE algorithm in detail. The evaluation of the effectiveness and efficiency of THUE is reported in Section \ref{sec:experiment}. Finally, the conclusions and future work are presented in Section \ref{sec:conclusions}.

\section{Related Work}
\label{sec:related}

In this section, we briefly review prior works about FEM, HUEM, and THUE mining, respectively.

\subsection{Frequent episode mining}

Research on FEM was first introduced by Mannila \textit{et al.} \cite{mannila1995discovering}, who adopted \textit{Apriori}-based discovery methods and defined many basic concepts of FEM. They found interesting frequent sub-sequences from alarm sequences in telecommunication networks by performing breadth-first searching (BFS). Note that FEM is different from the concept of sequential pattern mining (SPM) \cite{gan2019survey}. Mannila \textit{et al.} \cite{mannila1995discovering} proposed a sliding-window technique that mines parallel and serial episodes, where parallel means that a set of events occurs simultaneously, and a serial episode refers to events with alphabetical order. However, the support of occurrence may count more than once via a sliding window \cite{Iwanuma04onanti-monotone}. Another proposed core concept to circumvent this problem is minimal occurrences. Although \cite{mannila1995discovering} is a highlight achievement in the FEM domain, the problem of counting the occurrence of episodes repeatedly, and the shortcomings of the Apriori algorithm \cite{rathore2016top,wu2013mining} should be addressed. To solve these issues, EMMA \cite{huang2008efficient} utilizes depth-first search (DFS) and memory anchor techniques to mine frequent episodes.

DFS category algorithms find interesting patterns without generating candidates, but by expending prefix episodes in the sequence. Moreover, most of the aforementioned algorithms can only deal with simple event sequences, and few consider complex event sequences. However, complex event sequences are more realistic (e.g., customer transactions \cite{ao2017mining, fahed2018deer}, stock data \cite{lin2017novel}, cloud data \cite{amiri2018online}, etc.). In recent years, Ao \textit{et al.} \cite{ao2019large} focused on FEM research and proposed a scalable distributed framework for complex event sequences with event hierarchies. Owing to traditional episode rule mining algorithms that lack fine-grained response time, \cite{ao2017mining} proposed a solution to provide an ordered set of events, where the elapsed time between any two consecutive events is a constant. Other interesting results can be found in \cite{ao2015online,ao2018free}. DEER \cite{fahed2018deer} mines episode rules with a distant consequent (called \textit{distant episode rules}) and an antecedent as small as possible. It adopts support and temporal confidence metrics to filter out rules with a consequent that is close to the antecedent. Although many studies \cite{mannila1995discovering,huang2008efficient,wu2013mining,zimmermann2014understanding,ao2015online,ao2017mining,fahed2018deer,ao2018free,ao2019large,fournier2020tke} have been proposed for FEM, as discussed in the Introduction section, FEM algorithms may discover numerous patterns with low profit, and miss the highly profitable character of low-frequency patterns. In the next subsection, we introduce HUEM.

\subsection{High-utility episode mining}

In general, HUEM \cite{guo2012a,wu2013mining,guo2014high,lin2015discovering,gan2019utility,gan2019discovering,fournierviger2019hue-span} is significantly different than FEM and high-utility sequential pattern mining (HUSPM) \cite{gan2021survey,gan2020proum,gan2020fast,zhang2020tkus}. FEM algorithms always assume that all events are equally important, and this characteristic leads to the discovery of many episodes with low revenue but further reduces the low-frequency patterns having high revenue. In HUEM, the quantity of occurrence (i.e., internal utility) and the utility value of events (i.e., external utility) are taken into account. Guo \textit{et al.} \cite{guo2012a} first completed a study on HUEM. Nevertheless, they found HUEs from simple event sequences. Later, UP-Span \cite{wu2013mining} utilized two strategies, called Discarding Global unpromising Events (DGE) and Discarding Local unpromising Events (DLE), to prune low-utility events and reduce the searching cost in complex event sequences; in addition, as we described in introduction section, \textit{EWU} in HUEM is equivalent to \textit{TWU} in UPM. In practice, \textit{EWU} is not a special tight upper bound for episodes, as \textit{TWU} is. Generating candidates in complex event sequences is intricate, which causes UP-Span to suffer from a high running time and memory. Fortunately, TSpan \cite{guo2014high} enhanced UP-Span by adding a new prefix-tree structure and using two much tighter upper bounds (IEIC and IESC) to mine HUEs more efficiently.

However, both UP-Span and TSpan have the same drawback, caused by using a loose upper bound (\textit{EWU}) to prune the search space. Their output results may be incomplete and may even contain some low-utility episodes because of the episode-weighted downward closure property. UBER-Mine \cite{lin2015discovering} showed that \cite{wu2013mining} and \cite{guo2014high} without an appropriate anti-monotone property may produce rules with low utility. Lin \textit{et al.} \cite{lin2015discovering} firstly developed a straightforward method for developing HUE-based rules in the mining process from a complex event sequence. It is an updated version of the UP-Growth algorithm \cite{Tseng2010upgrowth}. First, UBER-Mine scans the complex event sequence and then records all 1-episode. Next, it finds simultaneous and serial events that are based on 1-episode. Finally, it uses all the episodes as input of UP-Growth and generate rules. Lin \textit{et al.} \cite{lin2015discovering} pointed out two highlights. First, a transactional database can also be seen as a complex event sequence, where each transaction ID is the same as an occurrence timestamp. Second, itemset pattern mining algorithms are potential solutions for episode mining tasks. The latest studies we found are UMEpi \cite{gan2019utility,gan2019discovering} and HUE-Span \cite{fournierviger2019hue-span}. The UMEpi algorithm is the first to discover complete HUEs without any loss or error. It redefines the computation method to filter unpromising episodes. HUE-Span utilizes a matrix structure to achieve simultaneous and serial concatenation operations. Its main contribution is integrating the concept of the remaining utility \cite{liu2012mining} into episode mining.

\subsection{Top-$k$ high-utility episode mining}

Similar to most UPM algorithms, the biggest challenge of HUEM is how to set an optimal threshold. The higher the minimum threshold (e.g., \textit{minUtil}, \textit{minSup}) we set, the less interesting patterns we obtain, and even some key information will be lost \cite{gan2020tophui}. On the contrary, if we set too low, the running time and memory consumption will be unacceptable because of the large number of candidates. It is a wholly troublesome task, which becomes a bottleneck for HUEM in large databases. The advantage of top-$k$ HUEM is that users no longer have to repeatedly test many thresholds. They simply set the number of HUEs they want, and the threshold is automatically raised. This is simple and convenient. Until now, few preliminary studies have been conducted to capture top-$k$ HUEs, and existing strategies require improvement in terms of runtime cost, memory consumption, unpromising candidate filtering, and scalability. The TUP algorithm \cite{rathore2016top} was the first to address this issue. It was concluded that the previous top-$k$ high-utility itemset mining algorithms cannot be applied in the top-$k$ HUEM directly in a huge complex event sequence. It uses two strategies, namely, the \textit{EWU} strategy and the pre-insertion strategy, to discover $k$ HUEs. The algorithm utilizes a DFS method to calculate new episodes and makes the current episode a prefix while exploring HUEs. Meanwhile, it uses a priority queue to maintain information about the current \textit{minUtil} threshold.

Based on the aforementioned preliminaries, we can draw the conclusion that there is little information available in the literature about finding HUEs, especially top-$k$ mining. Although the TUP algorithm is a breakthrough contribution, we can perform better than the TUP algorithm. Therefore, we develop a faster and more precise algorithm based on UMEpi \cite{gan2019utility,gan2019discovering} for mining the top-$k$ HUEM.

\section{Preliminaries and Problem Formulation}
\label{sec:preliminaries}

Based on previous studies, we primarily use notation and definitions given by \cite{gan2019utility,gan2019discovering} in this section. This paper focuses on the problem of discovering top-$k$ HUEs from complex event sequences. Additional details about HUEM can be found in \cite{rathore2016top, wu2013mining}.

\subsection{Preliminaries of HUEM}

\rm Each distinct event in a sequence is associated with an event type and occurrence time. An event is defined as a pair ($e_j$, $T_i$) that $e_j$ belongs to a finite alphabet $S$ = ($e_1$, $e_2$, $\ldots$, $e_{n-1}$, $e_n$) owns the symbolic event types, which can be defined in terms of weight, profit, risk or any other metric, depend on the user \cite{achar2012unified} \footnote{For the sake of simplicity, the notion ``event'' contains the ``event type'' concept in the remainder of this paper}. We call $T_i \in N^+$ the \textit{occurrence time point} when $e_j$ occurs. We assume that each event $e_j$ is associated with a positive number $p(e_j)$, namely, the external utility (e.g., profit and risk). Each event $e_j$ at the $T_i$ timestamp is associated with a positive number $q(e_j, T_i)$, called the internal utility (e.g., quantity). Through $p(e_j)$ $\times$ $q(e_j, T_i)$, we can obtain the utility $u(e_j, T_i)$ of each event.

\begin{figure}[htbp]
	\centering
	\includegraphics[scale=0.9]{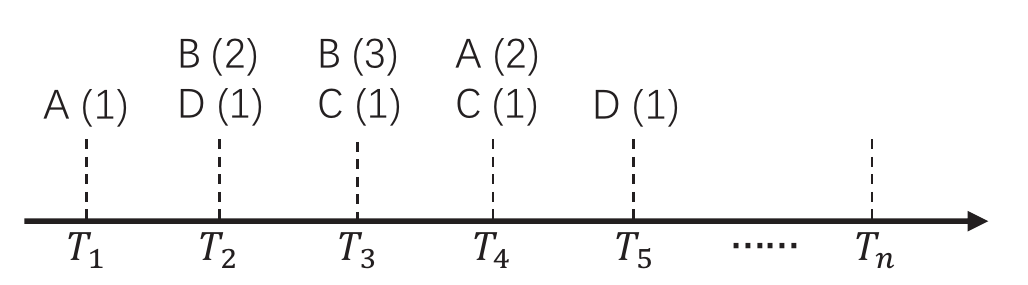}
	\caption{A simple complex event sequence.}
	\label{fig:data}	
\end{figure}

\begin{table}[htbp]
	\centering
	\caption{External utility value of events}
	\label{figUnitProfits}
	\begin{tabular}{|c|c|c|c|c|} \hline
		\textbf{Event} 	          & $A$ & $B$ & $C$ & $D$  \\ \hline
		\textbf{External Utility} & \$2 & \$1 & \$3 & \$2  \\ \hline
	\end{tabular}
\end{table}

\begin{definition}[Simultaneous event set]
	\rm A simultaneous event ($S\! The E_i$ $\subseteq$ $S$) set is composed of a finite set of events (can be different types), where each event occurrence has the same timestamp $T_i$ $\in$ $N^+$. We can see that ($\big<(BD)\big>$, $T_2$) in Fig. \ref{fig:data}. is a simultaneous event set that occurs at $T_2$. In particular, the event types in $S\!E_i$ also obey alphabetic order; for example, ($\big<(BD)\big>$, $T_2$) is not equivalent to ($\big<(DB)\big>$, $T_2$).
\end{definition}

\begin{definition}[Complex event sequence]
	\rm A \textit{complex event sequence} \textit{CES} = $\big<$($S\!E_1$, $T_1$), ($S\!E_2$, $T_2$), $\ldots$, ($S\!E_n$, $T_n$)$\big>$ consists of many simultaneous event sets in chronological order. In addition, each $S\!E_i$ is associated with unique time point $T_i$ ($i$ $\in$ $N^+$) and $T_i$ $<$ $T_j$ when 1 $\leq$ $i$ $<$ $j$ $\leq$ $n$.
\end{definition}

\begin{definition}[Episode]
	\rm An episode is a nonempty complex event sub-sequence that belongs to $S$. An $l$-episode means that the number of events is $\mid l \mid$, and the size of an episode is defined as the number of simultaneous event sets it contains. We call an episode that contains a simultaneous event set a complex episode, which does not include a simple episode \footnote{Generally, FEM often ignores complex episodes; for convenience, we use ``episode'' instead of ``complex episode'' and ``sample episode'' in the remainder of the paper}.
\end{definition}

\rm The length and size of episode $\big<$($A$) $\rightarrow$ ($BD$)$\big>$ are 3 and 2, respectively. Then, the simultaneous event set ($BD$) has a partial order (alphabetic or user-defined) $B$ $\succ$ $D$. Obviously, the size of all 1-episodes is also 1, such as $\big<$($A$)$\big>$, $\big<$($B$)$\big>$, $\big<$($C$)$\big>$.

\begin{definition}[Episode with simultaneous and serial concatenations]
	\rm  Let $\alpha$ = $\big<$$(E_1)$, $(E_2)$, $\ldots$, $(E_x)$$\big>$ and $\beta$ = $\big<$$(E_1^\prime)$, $(E_2^\prime)$, $\ldots$, $(E_y^\prime)$$\big>$ be two distinct episodes. The simultaneous concatenation of $\alpha$ and $\beta$ is defined as \textit{Simult-Concatenate}($\alpha$, $\beta$) = $\big<$$\alpha$ $\cup$ $\beta$$\big>$ = $\big<\big($$(E_1)$, $(E_2)$, $\ldots$, $(E_x)$, $(E_1^\prime)$, $(E_2^\prime)$, $\ldots$, $(E_y^\prime)$$\big)\big>$. The serial concatenation of $\alpha$ and $\beta$ is defined as \textit{Serial-Concatenate}($\alpha$, $\beta$) = $\big<$$\alpha$, $\beta$$\big>$ = $\big<$$(E_1)$, $(E_2)$, $\ldots$, $(E_x)$, $(E_1^\prime)$, $(E_2^\prime)$, $\ldots$, $(E_y^\prime)$$\big>$. Hence, simultaneous episodes are made by simultaneous concatenation, and serial concatenation constructs serial episodes.
\end{definition}

\rm Fig. \ref{fig:data} shows a complex event sequence \textit{CES} = $\big<$(($A$), $T_1$), (($BD$), $T_2$), (($BC$), $T_3$), (($AC$), $T_4$), (($D$), $T_5$)$\big>$. Among this sequence, $\big<$$A$$\big>$ and $\big<$$C$$\big>$ compose the simultaneous episode in $T_4$, while $\big<$($A$) $\rightarrow$ ($B$)$\big>$ is a serial episode. In particular, $\big<$$(A) \rightarrow (B)$$\big>$ and $\big<$$(B) \rightarrow (A)$$\big>$ are two distinct serial episodes since their sequence-orders are not the same.

\begin{definition}[Occurrence]
	\rm Set an episode $\alpha$ = $\big<$$(S\!E_1)$, $(S\!E_2)$, $\ldots$, $(S\!E_k)$$\big>$. The time interval [$T_s$, $T_e$] is called an occurrence of $\alpha$ iff: 1) $\alpha$ occurs in [$T_s$, $T_e$]; and 2) the first simultaneous event set $S\!E_1$ occurs at $T_s$ and the last simultaneous event set $S\!E_k$ occurs at $T_e$. The total occurrences of $S\!E_t$ ($1 \le t \le k$) of $\alpha$ in \textit{CES} is denoted by \textit{occSet}$(\alpha)$.
\end{definition}

\begin{definition}[Minimal occurrence]
	\rm Given two time intervals [$T_s$, $T_e$] and [$T_s^\prime$, $T_e^\prime$] are occurrences of an episode $\alpha$ separately, [$T_s^\prime$, $T_e^\prime$] is called sub-time interval of [$T_s, T_e$] iff $T_s$ $<$ $T_s^\prime$ and $T_e^\prime$ $<$ $T_e$. [$T_s$, $T_e$] is called a minimal occurrence of episode $\alpha$ iff: 1) [$T_s$, $T_e$] is an occurrence of $\alpha$; and 2) there is no occurrence [$T_s^\prime$, $T_e^\prime$] of $\alpha$ such that [$T_s^\prime$, $T_e^\prime$] is a sub-time interval of [$T_s$, $T_e$]. We let $mo$($\alpha$) denote the minimal occurrences of $\alpha$ with respect to [$T_s$, $T_e$]. The complete set of minimal occurrences of $\alpha$ in \textit{CES} is defined as \textit{moSet}($\alpha$) = $\{$$mo_1$($\alpha$), $mo_2$($\alpha$), $\ldots$, $mo_k$($\alpha$)$\}$.
\end{definition}

In Fig. \ref{fig:data}, the time interval [$T_2$, $T_3$] is a minimal occurrence of the episode $\big<$($BD$)$\big>$, and \textit{moSet}($\big<$($BD$)$\big>$) = \{[$T_2$, $T_3$], [$T_3$, $T_5$]\}. Therefore, we can easily know that the minimal occurrence is the shortest time interval that contains a particular episode.

\begin{definition}[Maximum time duration]
	\rm Maximum time duration (\textit{MTD}) is a user-specified time interval. An episode $\alpha$ satisfies the maximum time duration constraint iff ($T_e$ $-$ $T_s$) $\leq$ \textit{MTD}. Note that this inequality is different from UP-Span or TSpan, and we use this constraint to be easy to understand and more precise \cite{gan2019utility,gan2019discovering}. For convenience, we assume that \textit{mo}($\alpha$) always satisfies the maximum time duration constraint in this study.
\end{definition}

\begin{definition}[Sub-episode and super-episode]
	\rm Set two episodes $\alpha$ = $\big<$$S\!E_1$, $S\!E_2$, $\ldots$, $S\!E_n$$\big>$ and $\beta$ = $\big<$$S\!E_1^\prime$, $S\!E_2^\prime$, $\ldots$, $S\!E_m^\prime$$\big>$, where $m \leq n$, the episode $\beta$ is called a \textit{ sub-episode } of $\alpha$ iff there exist $m$ integers 1 $\leq$ $i_1$ $<$ $i_2$ $<$ $\ldots$ $<$ $i_m$ $\leq n$ such that $S\!E_{k}^\prime$ $\in$ \textit{CES} for 1 $\leq k \leq m$ $\leq n$. In contrast, episode $\alpha$ is called the \textit{super-episode} of $\beta$.
\end{definition}

For instance, the episode $\big<$($BC$) $\rightarrow$ ($D$)$\big>$ is called the \textit{ sub-episode } of $\big<$($B$), ($BC$), ($B$) $\rightarrow$ ($D$)$\big>$, whereas it is a \textit{super-episode} of $\big<$($C$) $\rightarrow$ ($D$)$\big>$.

\begin{definition}[Utility of a simultaneous event set w.r.t. a time point]
	\rm The utility of a simultaneous event set $S\!E_k$ $\subseteq$ \textit{CES} at timestamp $T_i$ is formulated as $u(S\!E_k, T_i)$ = $\sum_{e_j \in S\!E_k}$${u(e_j, T_i)}$. Hence, $u(S\!E_k, T_i)$ represents the total utilities generated by all events $e_j$ $\in$ $S\!E_k$ at the time point $T_i$.
\end{definition}

\rm In Fig. \ref{fig:data}, set \textit{SE}$_j$ = ($BD$), and $u(\big<(BD)\big>, T_2)$ = $u(B, T_2)$ + $u(D, T_2)$ = 2 $\times$ \$1  + 1 $\times$ \$2 = \$4.

\begin{definition}[Total utility of a complex event sequence]
	\rm Given a complex event sequence \textit{CES}, the total utility of all events in a timestamp $T_i$ is defined as $tu(T_i)$ = $\sum_{e_{j}\in T_i}$$u(e_{j}, T_i)$, where $e_j$ is the $j$-th event in $T_i$. Then, the total utility of \textit{CES}, denoted as \textit{TU}, is defined as \textit{TU} = $\sum_{T_i \in CES}$$tu(T_i)$.
\end{definition}

\rm Consider the third time point in Fig. \ref{fig:data}, $tu(T_3)$ = $u((B), T_3)$ + $u((C), T_3)$ = \$3 + \$3 = \$6. Then, the utilities of $T_1$ to $T_6$ can be calculated as \textit{tu}($T_1$) = \$2, \textit{tu}($T_2$) = \$4, \textit{tu}($T_3$) = \$6, \textit{tu}($T_4$) = \$7, and \textit{tu}($T_5$) = \$2. Therefore, the total utility in \textit{CES} is \textit{TU} = \$2 + \$4 + \$6 + \$7 + \$2 = \$21. To determine the utility of an episode, we first need to define how to count its occurrences. In the field of HUEM, we use the minimal occurrence to calculate the utility, which is described below.

\begin{definition}[Utility of an episode w.r.t. a minimal occurrence]
	\rm Let \textit{mo}($\alpha$) = [$T_s$, $T_e$] be the minimal occurrence of the episode $\alpha$ = $\big<$$(S\!E_1)$, $(S\!E_2)$, $\ldots$,  $(S\!E_k)$$\big>$, where each simultaneous event set $S\!E_j$ $\in$ $\alpha$ is associated with a distinct timestamp, $T_i$. Under the previous definition of minimal occurrence, the utility of episode $\alpha$ w.r.t. \textit{mo}($\alpha$) can be defined as $u(\alpha, mo(\alpha))$ = $\sum _{j=1}^{k}$$u(S\!E_j, T_i)$, $S\!E_j$ $\subseteq$ $\alpha$ and $T_s$ $\leq$ $T_i$ $\leq$ $T_e$.
\end{definition}

\rm Referring to the above example, assume $\alpha$ is $\big<$($B$) $\rightarrow$ ($C$)$\big>$. Because  \textit{moSet}($\alpha$) = \{[$T_2$, $T_3$], [$T_3$, $T_4$]\}, we have $u(\alpha, mo_{1}(\alpha))$ = $u((B), T_2)$ + $u((C)), T_3)$ = \$2 + \$3 = \$5, and $u(\alpha, mo_{2}(\alpha))$ = $u((B), T_3)$ + $u((C), T_4)$ = \$3 + \$3 = \$6.

\begin{definition}[Utility of an episode w.r.t. an event sequence]
	\rm Consider the entire event sequence \textit{CES}. Let $u(\alpha)$ denote the total utility of an episode $\alpha$ in \textit{CES} and \textit{moSet}($\alpha$) = [$mo_1(\alpha)$, $mo_2(\alpha)$, $\ldots$, $mo_n(\alpha)$] denote the complete set of minimal occurrences of $\alpha$ in \textit{CES}. Then $u(\alpha)$ = $\sum_{i=1}^{n}$$u(\alpha, mo_i(\alpha))$.
\end{definition}

\begin{definition}[High-utility episode]
	\rm Given a complex event sequence \textit{CES}, the maximum time duration \textit{MTD}, and a user-specify minimum utility (\textit{minUtil}) threshold, an episode is supposed as a high-utility episode (HUE) iff its utility in \textit{CES} is no less than \textit{minUtil} $\times$ \textit{TU} \cite{gan2019utility,gan2019discovering}. Otherwise, we call it a low-utility episode (LUE).
\end{definition}

\rm In Fig. \ref{fig:data}, the utility of episode $\alpha$ = $\big<$($B$) $\rightarrow$ ($AC$)$\big>$ is $u$($\alpha$) = \$4 + \$6 = \$10. If we assume \textit{MTD} = 2 and \textit{minUtil} = 45\%, $\alpha$ is an HUE, but episode $\big<$($BD$) $\rightarrow$ ($B$)$\big>$ is less than 45\% $\times$ \$21 (= \$9.45); thus, it is an LUE. Assuming \textit{MTD} = 2 and \textit{minUtil} = 45\%, the complete set of HUEs in Fig. \ref{fig:data} is shown in Table \ref{table:HUEs}.

Comparing the second and third columns in Table \ref{table:HUEs}, there are distinct differences between FEM and HUEM. In view of the frequency dimension, all the patterns have the same support. If we suppose that the minimal support is 2, then all episodes (except the second row) in the table should be deleted. Then, all 1-episodes are frequent patterns absolutely. However, their real utilities are less than the threshold (\$9.45). Obviously, the discovered frequent patterns may bring low utility (e.g., profit) because the support measure cannot correctly reflect the potential value of goods. Obviously, the results represent the limitation of the frequency metric. In particular, the utility and frequency of the third row are not \$19 and 2, respectively, because of the minimal occurrence concept. [$T_3$, $T_4$] is a sub-time interval of [$T_2$, $T_4$]. Thus, we obtain \$10 (utility) and 1 (frequency) in [$T_3$, $T_4$].

\begin{table}[htbp]
	\centering
	\caption{Final high-utility episodes} 
	\label{table:HUEs}
	\begin{tabular}{|c|c|c|} \hline
		\textbf{Episode}   &   \textbf{Utility}	&	\textbf{Frequency}      \\ \hline
		$\big<$($BC$) $\rightarrow$ ($AC$)$\big>$ & \$13 & 1 \\ \hline
		$\big<$($B$)  $\rightarrow$ ($C$)$\big>$  & \$11 & 2 \\ \hline
		$\big<$($B$)  $\rightarrow$ ($AC$)$\big>$ & \$10 & 1 \\ \hline
		$\big<$($BC$) $\rightarrow$ ($A$)$\big>$  & \$10 & 1 \\ \hline
		$\big<$($BD$) $\rightarrow$ ($BC$)$\big>$ & \$10 & 1 \\ \hline
		$\big<$($C$)  $\rightarrow$ ($AC$)$\big>$ & \$10 & 1 \\ \hline
		$\big<$($A$)  $\rightarrow$ ($C$)$\big>$  & \$10 & 1 \\ \hline
	\end{tabular}
\end{table}

\begin{definition}[Top-$k$ high-utility episode]
	\rm Sort all HUEs with decrement order of utility value, and get the top-$k$ HUEs as output. The parameter $k$ represents the number of HUEs users want, and users no longer need to test the algorithm again and again to find a suitable threshold. If we set $k$ = 2, \textit{MTD} = 2, we finally obtain two HUEs from Table \ref{table:HUEs}, $\big<$($BC$) $\rightarrow$ ($AC$)$\big>$, $\big<$($B$) $\rightarrow$ ($C$)$\big>$.
\end{definition}

\rm Assume that we set the initial $k$ = 4, \textit{MTD} = 3, and \textit{minUtil} = 0. After calculation, we can obtain the top-4 highest utility episodes: $\big<$($BD$) $\rightarrow$ ($BC$) $\rightarrow$ ($AC$)$\big>$ (= \$17), $\big<$($D$) $\rightarrow$ ($BC$)$\big>$ (= \$15), $\big<$($B$) $\rightarrow$ ($BC$) $\rightarrow$ ($AC$)$\big>$ (= \$15), and $\big<$($BC$) $\rightarrow$ ($AC$) $\rightarrow$ ($D$)$\big>$ (\$15). Thus, the current \textit{minUtil} is \$15.

\subsection{Problem Statement}
\label{def_HUEM}

In summary, the HUEM algorithm works on a long event sequence with timestamp characteristics. HUEM is different from FEM and high-utility sequential pattern mining (HUSPM) because both the problem formulation and mining mechanism are not the same. Based on the previous definitions, we can obtain the problem formulation of the top-$k$-based HUEM as follows:

\rm \textbf{Problem statement}: Given a complex event sequence with simultaneous or serial events that have external and internal utility, take the \textit{MTD} as the constraint, and a user-specified \textit{minUtil}. The main task is to solve the problem of top-$k$ HUE mining. In other words, it aims to mine the highest $k$ episodes whose utilities are no less than \textit{minUtil} $\times$ \textit{TU} as the required constraint \cite{gan2019utility,gan2019discovering}.

Traditional FEM algorithms do not consider the internal utility (e.g., purchase quantities) and external utility (e.g., unit profit) information of items. Therefore, FEM algorithms find frequent episodes and discard vital information. As a result, many uninteresting frequent patterns with low external utility may be discovered, and some episodes generating high profits may be missed. However, HUEM algorithms consider the case where episodes have a weight/utility (i.e., importance, interest, or risk) and can occur more than once in a timestamp; for example, a rare jewel brings high profit but low sales volume; a pencil is the exact opposite case. To date, HUEM algorithms have gained immense importance owing to their multiple applications (e.g., different web pages).

\section{Proposed THUE Algorithm}
\label{sec:algorithms}

In this section, we propose an efficient algorithm, THUE, to discover top-$k$ high-utility episodes that satisfy the constraints of \textit{MTD} and \textit{minUtil}. THUE discovers HUEs by spanning the search space with respect to a conceptual lexicographic sequence tree (LS-tree). Moreover, the remaining utility of episodes and a tight upper bound are developed and utilized in some pruning strategies. Details of the downward closure property of \textit{EWU}, the pruning strategies with optimized \textit{EWU}, and the main procedures of THUE are described below.

\subsection{Downward Closure Property Pruning Strategy}

\begin{definition}[$I$-Concatenation and $S$-Concatenation]
	\rm Given an $l$-episode $\alpha$, when an event is appended to the end of $\alpha$, it will get a new episode. This process is called \textit{concatenation}. More specifically, if the duration time of the new extended episode is the same as the original episode, we call this operation $I$-concatenation. On the contrary, if the time duration is increased, we call this operation $S$-concatenation.
\end{definition}

\rm Consider a 1-episode $\alpha$ = $\big<$($B$)$\big>$ at the time point $T_2$, the concatenation set $\big<$($BD$)$\big>$ is its $I$-Concatenation, while $\big<$($B$) $\rightarrow$ ($D$)$\big>$ is the result of $S$-Concatenation process.

\begin{definition}[Lexicographic sequence tree]
	\rm In the lexicographic sequence tree \cite{ayres2002sequential}, a) the root node of the prefix-based tree is empty; b) for a parent node (also called a prefix node, represents an episode), all the child nodes are generated according to $I$-Concatenation and $S$-Concatenation operations; and c) all the child nodes of a prefix node are listed in a specific order (e.g., incremental order, arbitrary order, or lexicographic order).
\end{definition}

Given a long complex event sequence, the LS-tree is a structure that contains all the candidate episode information. In particular, the number of all possible episodes in the search space is extremely large \cite{laxman2005discovering,tatti2011mining}. Obviously, performing an exhaustive search (e.g., enumerate and then determine all possible episodes) is not an acceptable solution. Because many datasets always contain huge piles of low-utility patterns, filtering all high-utility patterns is not an easy task. Most existing studies have demonstrated that the utility metric is neither monotonic nor anti-monotonic \cite{liu2005two}. Based on the previous introduction of HUEs, we know that they do not hold the anti-monotonicity property. In other words, an episode may have a lower, equal, or higher utility than any of its sub-episodes. Without anti-monotonicity as the pruning constraint, it is difficult to efficiently reduce the search space in HUEM.

In the HUEM field, the concept of episode-weighted utilization (\textit{EWU}) \cite{guo2014high} was proposed as a utility upper bound of an episode in \textit{CES}. Although this work achieves a speed-up of several orders of magnitude over an exhaustive search, \textit{EWU} is still a loose upper bound \cite{gan2019discovering,fournierviger2019hue-span}. Thus, the \textit{EWU} value in existing algorithms is not complete, and previous studies on HUEM cannot extract all the HUEs. Next, we recommend a tighter upper bound that is slightly different from the original \textit{EWU}.

\begin{definition}[Episode-Weighted Utilization \cite{guo2014high}]
	\rm Set $mo(\alpha)$ = $[T_s, T_e]$ a minimal occurrence of the episode $\alpha$ = $\big<(E_1$), $(E_2)$, $\ldots$, $(E_k)\big>$, where each simultaneous event set $S\!E_i \in \alpha $ is associated with the time point $T_i$ (1 $\leq i$ $\leq k$), and $mo(\alpha)$ satisfies \textit{MTD}. The episode-weighted utilization of $\alpha$ w.r.t. $mo(\alpha)$ consists of two parts: 1) the utility of $\alpha$ and 2) the utility of extended candidate event sets. Its formula is \textit{EWU}($\alpha$, $mo(\alpha)$) = $\sum _{i=1}^{k}u(E_i, T_i)$ + $\sum _{i=T_{e}}^{T_{s} + MTD}u(S\!E_i, T_i)$, where $S\!E_i$ is a simultaneous event set that occurs at $T_i$ in \textit{CES}. Apparently, \textit{EWU} w.r.t. $moSet$($\alpha$) = [$mo_{1}(\alpha)$, $mo_{2}(\alpha)$, $\ldots$, $mo_{n}(\alpha)$], such that \textit{EWU}($\alpha$) = $\sum _{i=1}^{n}EWU(\alpha, mo_{i}(\alpha))$. More details can be found in \cite{gan2019discovering}.
\end{definition}

\rm Given an \textit{MTD} = 2, the \textit{EWU} of $\alpha$ = $\big<(BD)$ $\rightarrow$ $(C)\big>$ w.r.t. its $mo(\alpha)$ = $[T_2, T_3]$ is calculated as $EWU(\alpha, [T_2, T_3])$ = $\big(u((BD), T_2)$ + $u((C), T_3)\big)$ + $u(\big<(BC)\big>, T_3)$ + $u(\big<(AC)\big>, T_4)$ + $u(\big<(D)\big>)$ = $\big($\$4 + \$3$\big)$ + \$6 + \$7 + \$2 = \$22.

\begin{definition}[High Weighted Utilization Episode \cite{guo2014high}]
	\rm Given a complex event sequence \textit{CES}, an episode is called a high weighted utilization episode (HWUE) in \textit{CES} iff its \textit{EWU} is no less than \textit{minUtil} $\times$ \textit{TU}. HWUE is also called a promising episode. This indicates that it may be a potential HUE. Otherwise, it is an unpromising episode, and we can ignore it and its super-episodes directly. The theorem and details of the proof can be found in \cite{gan2019discovering}.
\end{definition}

\begin{strategy}[Episode-Weighted Downward Closure Pruning Strategy]
	\label{strategy:EWU}
	\rm Let $\alpha$ and $\beta$ be two different episodes, and $\gamma$ is a super-episode of $\alpha$ and $\beta$, either generated by \textit{Simult-Concatenate}($\alpha$, $\beta$) or \textit{Serial-Concatenate}($\alpha$, $\beta$). The episode-weighted downward closure (EWDC) property means that if \textit{EWU}($\alpha$) $<$ \textit{minUtil} $\times$ \textit{TU} or \textit{EWU}($\beta$) $<$ \textit{minUtil} $\times$ \textit{TU}, $\gamma$ must be a low utility episode. It is important to note that \textit{EWU} is the upper bound of the episode as a prefix when performing prefix spanning. In other words, the non-HWUE may still be the sub-episode of the final HUEs (as the suffix in HUEs). Thus, we cannot remove those episodes that are not HWUEs when prefix-spanning is used to discover HUEs. Readers can refer to \cite{guo2014high,gan2019discovering,fournierviger2019hue-span} for the proof.
\end{strategy}

Based on the above definitions, we obtain two vital observations. Here, we only list the conclusions; the details can be found in \cite{gan2019discovering}. First, \textit{EWU} can utilize the downward closure property as a loose upper bound of HUEM, but it is distinct from \textit{TWU} or \textit{SWU}. \textit{EWU} just works in the subtree of an LS-tree instead a global tree. Second, according to \textit{EWU} as a single upper bound, we may discover an incomplete set of HUEs.

\subsection{Remaining Utility Pruning Strategy}

As previously mentioned, an episode mining algorithm that only utilizes the EWDC property to reduce the search space is not a good method. Because the previous works compute \textit{EWU}($\alpha$) with respect to $mo(\alpha)$, it is divided into two parts: the utility value of episode $\alpha$ in the $mo(\alpha)$ time point, and the remaining episodes of $\alpha$ starting from time point $T_{\rm e}$ to $T_{\rm s}$ + \textit{MTD}. Furthermore, if we anatomize the derivation process of this formula, we can find that the episode utility of $\alpha$ in the $T_{\rm e}$ occurrence time point has a computed overlap. This explains why the original \textit{EWU} is a loose upper bound. To improve the mining performance, Gan \textit{et al.} \cite{gan2019discovering} proposed two optimized strategies to decrease the upper bound of \textit{EWU}. In the next section, we discuss how to optimize \textit{EWU} and describe the corresponding pruning strategies.

\begin{definition}[Remaining Utility of an Episode \cite{gan2019discovering}]
	\rm Given a time point $T_i$, the remaining utility of an episode $\alpha$ in $T_i$ is the accumulative utilities of the remaining events after this episode in $T_i$: $ru(\alpha, T_i)$ = $\sum_{e^\prime\notin \alpha \wedge \alpha \prec e^\prime}$$u(e^\prime, T_i)$.
\end{definition}

\rm Let $\alpha$ = $\{A\}$ be in $T_4$; the remaining utility of $\alpha$ is $ru(\alpha, T_4)$ = $u(\{C\}, T_3)$ = \$3. Obviously, the remaining utility is \$0 when the remaining event set of $\alpha$ is empty, such as $ru(\{A\}, T_1)$ = \$0.

\begin{table}[htbp]
	\centering
	\caption{Comparison of \textit{EWU} and \textit{EWU}$_{\rm opt}$}
	\label{table:Comparison}
	\begin{tabular}{|c|c|c|} \hline
		\textbf{1-Episode} & \textbf{\textit{EWU}} & \textbf{\textit{EWU}$_{\rm{opt}}$} \\ \hline
		$A$ & \$27  & \$21  \\ \hline
		$B$ & \$37  & \$32  \\ \hline
		$C$ & \$30  & \$24  \\ \hline
		$D$ & \$23  & \$19  \\ \hline
	\end{tabular}
\end{table}

\begin{definition}[Optimized Episode-Weighted Utilization \cite{gan2019discovering}]
	\rm The optimized episode-weighted utilization (\textit{EWU}$_{opt}$) of $\alpha$ = $\big<$($S\!E_1$), ($S\!E_2$), $\ldots$, ($S\!E_{k-1}$), ($S\!E_k$)$\big>$ w.r.t. $mo(\alpha)$ consists of three parts: 1) the utility of $\alpha$, 2) the remaining utility of $\alpha$ in $T_{e}$, and 3) the utility of all episodes in [$T_{e}$+1, $T_{s}$+\textit{MTD}]. Then, the optimized episode-weighted utilization of $alpha$ is defined as \textit{EWU}$_{opt}$($\alpha$, $mo(\alpha)$) = $u(\alpha, mo(\alpha))$ + $ru(S\!E_k, T_e)$ + $\sum _{i=T_{e}+1}^{T_{s} + MTD}tu(T_i)$ = $\sum _{i=1}^{k}u(S\!E_i, T_i)$ + $ru(S\!E_k, T_e)$ + $\sum _{i=T_{e}+1}^{T_{s} + MTD}tu(T_i)$, where the timestamp $T_i$ is within the satisfied \textit{MTD} interval $[T_e, T_{s}$+\textit{MTD}]. Thus, for $\alpha$ in \textit{CES}, we easily obtain the accumulative \textit{EWU} w.r.t. \textit{moSet}($\alpha$) = [$mo_{1}, mo_{2}, \ldots, mo_{n}$], such that \textit{EWU}($\alpha$) = $\sum _{i=1}^{n}$\textit{EWU}($\alpha, mo_n(\alpha)$).
\end{definition}

\rm Set \textit{MTD} = 2; Table \ref{table:Comparison} shows the difference between \textit{EWU} and \textit{EWU}$_{opt}$ for 1-episodes. Clearly, the original \textit{EWU} is greater than \textit{EWU}$_{opt}$. It can be seen that $\alpha = \big<A\big>$; it occurs at timestamp [$T_1, T_1$] and [$T_4, T_4$], while computing the original \textit{EWU}, so that the utility value of event $\{A\}$ is double-counted. We can obtain the derivation process \textit{EWU}$(\alpha)$ - \textit{EWU}$_{opt}(\alpha)$ = \$27 - \$21 = \$6 = \$2 + \$4 = $u(\alpha, T_1)$ + $u(\alpha, T_4)$.

\begin{table}[htbp]
	\centering
	\caption{Results of each 1-episode}
	\label{table:EWU}
	\begin{tabular}{|c|c|c|c|} \hline
		\textbf{1-Episode} & \textbf{\textit{moSet}} & \textbf{\textit{EWU}} & \textbf{Utility} \\ \hline
		$A$ & \{[$T_1,T_1$], [$T_4,T_4$]\} & \$21  & \$6  \\ \hline
		$B$ & \{[$T_2,T_2$], [$T_3,T_3$]\} & \$32  & \$5  \\ \hline
		$C$ & \{[$T_3,T_3$], [$T_4,T_4$]\} & \$24  & \$6  \\ \hline
		$D$ & \{[$T_2,T_2$], [$T_5,T_5$]\} & \$19  & \$4  \\ \hline
	\end{tabular}
\end{table}

Specifically, the \textit{EWU}\footnote{If not otherwise specified, \textit{EWU} represents \textit{EWU}$_{opt}$ in the explanation that follows.} value of $\alpha$ in \textit{CES} is always greater than or equal to the total utility of $\alpha$, as well as the total utility of its super-episodes in the search space. The results of each 1-episode and its \textit{moSet}, \textit{EWU}, and real utility values are listed in Table \ref{table:EWU}. With the conceptual LS-tree, the upper bound \textit{EWU} has the \textit{local downward closure} property. Based on previous observations, we can use the following filtering strategies. First, strategy \ref{strategy:EWU} provides a way to track patterns that may be HUEs in the subtree. This guarantees the return of exact HUEs. Therefore, it is a vital property that our novel algorithm utilizes the tighter upper bound \textit{EWU}, as we describe below.

\begin{strategy}[Optimized \textit{EWU} pruning strategy  \cite{gan2019discovering}]
	\rm When spanning the LS-tree rooted at an episode $\alpha$ as the prefix, the proposed algorithm spans/explores the search space in a depth-first search way. If the \textit{EWU} of any node/episode $\alpha$ is less than \textit{minUtil} $\times TU$, then any of its child nodes would not be a final HUE; they can be regarded as low-utility episodes and cut directly.
\end{strategy}

The conceptual LS-tree is a prefix-shared tree, and the algorithm extends the lower-level nodes/episodes to the higher-level ones in the depth-first search method. Strategy \ref{strategy:EWU} first establishes a theoretical basis for generating high-utility $l$episodes ($l$ $\geq$ 2) from the 1-HWUEs. The basic idea of THUE is to find all 1-HWUEs by calculating the \textit{EWU} values of all 1-episodes, and then directly extending $l$-episodes ($l$ $\geq$ 2). Therefore, THUE discovers the $l$-HWUEs and $l$-HUEs by extending each 1-HWUE as a prefix. For any $k$-episode node (k $\geq$ 2) in the LS-tree, we can easily obtain its 1-extensions using $I$-concatenation and $S$-concatenation.

Specifically, Strategy \ref{strategy:EWU} provides the necessary conditions for computing all prefix-based HWUEs and HUEs. \textit{EWU} is the upper bound on utility for any extension of a node/episode in a subtree, and when any node/episode has a low upper bound, such as \textit{EWU}($\alpha$) $<$ \textit{minUtil} $\times$ \textit{TU}, we assume that they are not HWUEs and cannot be $\alpha$-prefixed episodes. Thus, we avoid searching for unpromising nodes and further save runtime.

\subsection{Raising Threshold Strategies}

To the best of our knowledge, we find that only two algorithms, TKE \cite{fournier2020tke} and TUP \cite{rathore2016top}, are related to the field of top-$k$ episode mining. They discovered frequent episodes and HUEs, respectively. Their designed raising threshold strategies are both effective, but not very efficient. It should be noted that we can review some increasing \textit{minUtil} techniques used in top-$k$ high-utility itemset mining from \cite{krishnamoorthy2019comparative}. In the following section, we introduce the threshold strategies adopted in our novel algorithm.

\begin{table}[htbp]
	\centering
	\caption{The real utility of each event}
	\label{table:realUtil}
	\begin{tabular}{|c|c|c|c|c|} \hline
		\textbf{Event} 	      & $A$ & $B$ & $C$ & $D$  \\ \hline
		\textbf{Real Utility} & \$6 & \$5 & \$6 & \$4  \\ \hline
	\end{tabular}
\end{table}

\begin{strategy}[\textit{RIU}: Raising threshold based on the real event utilities \cite{ryang2015rept}]
	\label{strategy:riu}
	\rm After dataset scanning, RIU is applied to raise the threshold. This strategy increases the current \textit{minUtil} using real event utilities. While scanning the dataset, RIU uses a hash map or array to compute each event utility for each simultaneous event set. Therefore, we can obtain the real utility of each event. Table \ref{table:realUtil} is a real-event utility table for events shown in Fig. \ref{fig:data}.
\end{strategy}

Obliviously, sometimes only adopting real event utilities is insufficient. From Table \ref{table:realUtil}, we can see that if we set $k$ = 4, the raising threshold is just \$4. In the previous discussion, the final \textit{minUtil} is \$15 when $k$ = 4, \textit{MTD} = 3 from Fig. \ref{fig:data}. The current \textit{minUtil} is much lower than that of the final \textit{minUtil}. Therefore, we utilize another raising strategy (called \textit{RTU}) after the \textit{RIU} method.

\begin{table}[htbp]
	\centering
	\caption{The real utility of each event}
	\label{table:tUtil}
	\begin{tabular}{|c|c|c|c|c|c|} \hline
		\textbf{Timestamp} & $T_1$ & $T_2$ & $T_3$ & $T_4$ & $T_5$ \\ \hline
		\textbf{TU} 	   & \$2   & \$4   & \$6   & \$7   & \$2   \\ \hline
	\end{tabular}
\end{table}

\begin{strategy}[\textit{RTU}: Raising threshold based on transaction utilities \cite{gan2020tophui}]
	\label{strategy:rtu}
	\rm After dataset scanning, \textit{RTU} also uses a hash map or array to compute each simultaneous event set (transaction). Then, the $k$-{th} highest \textit{TU} value is the latest \textit{minUtil} (if the current threshold is lower) until other strategies increase the threshold. Table \ref{table:tUtil} is a real event utility table for events from Fig. \ref{fig:data}.
\end{strategy}

For example, after combining the contents of Tables \ref{table:realUtil} and \ref{table:tUtil}, we obtain a utility list that is sorted from largest to smallest ($\{$\$7, \$6, \$6, \$6, \$5, \$4, \$4, \$2, \$2$\}$). Because $k$ = 4, we set the initial \textit{minUtil} as \$6, which is larger than \$4. If the range of utility values of events in long complex event sequences is large, the advantage of threshold-raising strategies will be more obvious.

\begin{strategy}[\textit{RUC}: Raising threshold by utility of candidates \cite{tseng2015tku-tko}]
	\label{strategy:ruc}
	\rm This strategy can be incorporated with any utility pattern mining algorithm. When dealing with HUEM, RUC utilizes a priority queue structure \textit{topKBuffer} to maintain the top-$k$ HUEs, where episodes are sorted in descending order of utility. Initially, \textit{topKBuffer} is an empty priority queue. Then, during the $I$-concatenation and $S$-concatenation search procedure, if the utility of an episode $\alpha$ is no less than \textit{minUtil}$_{current}$, $\alpha$ is added to \textit{topKBuffer}. Therefore, the threshold can be safely raised to the $k$-{th} highest utility in \textit{topKBuffer}. Subsequently, the episodes in which utilities are less than \textit{minUtil}$_{current}$ should be removed from \textit{topKBuffer}. Finally, we obtain the $k$ highest HUE list without any loss.
\end{strategy}

To give a simple example, we set $k$ = 2, and the current \textit{topKBuffer} list and \textit{minUtil} are $\{$$B$(\$5), $D$(\$4)$\}$, and \$4, respectively. Then, the process calculates a new episode $\{A(\$6)\}$, whose utility is higher than the current threshold. Therefore, the update \textit{topKBuffer} list is $\{$$A$(\$6), $B$(\$5)$\}$, and \textit{minUtil}$_{current}$ is also \$5. Naturally, $\{D(\$4)\}$ becomes a low-utility episode, and we delete it from the list.

\subsection{Mining Procedure}

After illustrating the lexicographic sequence tree, the optimized \textit{EWU}$_{opt}$ pruning strategy and other basic concepts of high utility episode mining were addressed. We now discuss the proposed algorithm in this subsection. To summarize, the main procedure of the THUE algorithm is described in \textbf{Algorithm \ref{AlgorithmTHUE}}. It takes a complex sequence \textit{CES}, \textit{MTD} and the number of users need $k$ as input parameters. We divide the proposed algorithm into two parts. 1) In Phase I, THUE mainly prepares for Phase II, initialized \textit{minUtil} as 0 (Line 1). Next, it scans \textit{CES} to construct the transformed event sequence $S^\prime$ in memory, and all the necessary mining information needed afterward has been stored. At the same time, we input the utility of each transaction and item as utility list parameters, and then call \textbf{Algorithm \ref{AlgorithmRUS}} to raise \textit{minUtil} (Lines 2--4). After obtaining all 1-episode candidates, Phase I has been completed (Lines 5--6). 2) In Phase II, THUE uses the depth-first search method and recursively calls the \textit{Span-SimultHUE} (\textbf{Algorithm \ref{Simul-search}}) and \textit{Span-SerialHUE} (\textbf{Algorithm \ref{Serial-search}}) procedures to discover the set of $l$-HWUEs and $l$-HUEs with $\alpha$ as a prefix (Lines 7--11). Finally, the $k$ highest utility episodes are output.

The THUE algorithm adopts Strategies \ref{strategy:riu} and \ref{strategy:rtu} called \textbf{R}aising \textbf{U} utility \textbf{S}trategy method (\textbf{Algorithm \ref{AlgorithmRUS}}) to automatically increase the \textit{minUtil}. It requires a list \textit{uList} storing candidate values to update \textit{minUtil} and parameter $k$, which is the desired number of episodes. Each element in \textit{uList} must be sorted because it always takes the $k$-{th} highest utility as the latest \textit{minUtil} (Line 1). Lines 2--6 aim to solve two different cases. 1) There are already some elements in list, and the quantity is no less than $k$. Then, the utility of the $k$-{th} value is chosen as the current \textit{minUtil} (Line 3). 2) In the second case, the list is empty at the beginning or the number of elements is less than $k$. Set the utility of the last element in \textit{uList} as the initialized \textit{minUtil} (Line 5). Finally, the procedure outputs the current minimal utility value to help prune the latter searching space process. Further, \textbf{Algorithm \ref{AlgorithmRUC}} utilizes Strategy \ref{strategy:ruc} to raise the threshold when mining $l$episodes. It always keeps a priority queue \textit{top-k HUEs} dynamically. When a new HUE is discovered, it can be saved into the set of \textit{top-k HUEs} immediately (Line 1). If the size of the \textit{top-k HUEs} is no less than $k$, then we sort the queue in decreasing order. Because it may occur that $k$-1 utilities are greater than the $k$-{th}, after inputting a new utility element, we should change the \textit{minUtil}$_{current}$ (Line 4). Finally, Line 5 removes low-utility episodes in \textit{top-k HUEs}.

\begin{algorithm}
	\caption{The THUE algorithm}
	\label{AlgorithmTHUE}
	\begin{algorithmic}[1]	
		\Require \textit{CES}: a complex event sequence; \textit{MTD}: user-specified maximum time duration; \textit{k}: desired number of episodes.
		\Ensure \textit{top-k HUEs}: a total top-$k$ high utility episodes.
		
		\State \textit{minUtil} $\leftarrow$ 0;
		
		\State scan \textit{CES} to get the transformed $S^\prime$ firstly, collect utility of each transaction and each 1-episode, store in list \textit{rtu} and \textit{riu} respectively;
		
		\State call \textbf{RUS}\textbf{(\textit{rtu}, $k$)};
		
		\State call \textbf{RUS}\textbf{(\textit{riu}, $k$)};
		
		\State calculate the original \textit{EWU};	
		
		\State sort all 1-episodes with the total order $\succ$;
		
		\For{each 1-episode $ e \in S^\prime$ under $\succ$ order}
		\If{ \textit{EWU}$(e) \geq minUtil$ }
		
		\State call \textbf{Span-SimultHUE}\textbf{($e$, $S^\prime$, \textit{MTD}, \textit{k})};
		
		\State call \textbf{Span-SerialHUE}\textbf{($e$, $S^\prime$, \textit{MTD}, \textit{k})};
		\EndIf
		
		\EndFor 		 					
		
		\State \textbf{return} \textit{top-k HUEs}
	\end{algorithmic}
\end{algorithm}

\begin{algorithm}
	\caption{The RUS procedure}
	\label{AlgorithmRUS}
	
	\begin{algorithmic}[1]
		\Require \textit{uList}: a utility value list; \textit{k}: desired number of episodes.
		\Ensure \textit{minUtil}: the current minimal utility.
		
		\State sort \textit{uList} list in numerical order;
		\If {the size of \textit{uList} $\ge$ \textit{k}}
		
		\State \textit{minUtil} $\leftarrow$ the utility of $k$-{th} highest episode;
		
		\Else
		\State \textit{minUtil} $\leftarrow$ the utility of last one in \textit{uList};
		
		\EndIf
		
		\State \textbf{return} a novel \textit{minUtil}
		
	\end{algorithmic}
\end{algorithm}

\begin{algorithm}
	\caption{The RUC procedure}
	\label{AlgorithmRUC}
	
	\begin{algorithmic}[1]
		\Require $\beta$: a HUE; \textit{k}: desired number of episodes.
		
		\State \textit{top-k HUEs} $\leftarrow$ $\beta$;
		
		\If {$\mid$\textit{top-k HUEs}$\mid$ $\ge$ $k$}
		\State sort appended \textit{top-k HUEs} list;
		\State \textit{minUtil} $\leftarrow$ utility of the $k$-{th} highest episode in \textit{top-k HUEs};
		
		\State remove the low-utility episodes in \textit{top-k HUEs};
		\EndIf
		
	\end{algorithmic}
\end{algorithm}

\textbf{Algorithm \ref{Simul-search}} and \textbf{Algorithm \ref{Serial-search}} show the \textit{Span-SimultHUE} and \textit{Span-SerialHUE} procedures, respectively. Both require the following parameters: 1) a prefix episode $\alpha$; 2) the transformed event sequence $S^\prime$; 3) \textit{MTD}; and 4) $k$. However, these two algorithms differ slightly. $I$-Concatenation is the key idea of the \textit{Span-SimultHUE} procedure, and this operation does not change the original episode $\alpha$ length. It only considers those events that occur in the same timestamp of $\alpha$, and the work steps are as follows. It first initializes the \textit{simultEpiSet} list as an empty set (Line 1), then obtains all simultaneous events of $\alpha$ based on $S^\prime$ and stores it in the \textit{simultEpiSet} list. Note that these appended simultaneous events follow the $\succ$ order (Lines 2--4). Subsequently, we iterate the elements in the \textit{simultEpiSet} list, call the function \textit{Simult-Concatenate($\alpha$, e}) to construct a new simultaneous episode $\beta$, and compute its \textit{EWU} value based on $S^\prime$ (Lines 5--7). Later, it is the same as \textbf{Algorithm \ref{AlgorithmTHUE}}, if the utility of the simultaneous episode $\beta$ is no less than the current \textit{minUtil}, it takes a new episode $\beta$ into the \textit{top-k HUEs} list and updates the current \textit{minUtil} if necessary (Lines 9--14).

\begin{algorithm}
	\caption{The Span-SimultHUE procedure}
	\label{Simul-search}
	
	\begin{algorithmic}[1]	
		\Require $\alpha$: the current episode; $S^\prime$: the transformed event sequence; \textit{MTD}: user-specified maximum time duration; \textit{k}: desired number of episodes.
		\Ensure \textit{top-k HUEs}: a set of top-$k$ high-utility episodes taking $\alpha$ as prefix.
		
		\State initialize \textit{simultEpiSet} = $\phi$;
		
		\For{each \textit{mo}($\alpha)$ = $[T_s, T_e] \in$ \textit{moSet}($\alpha$)}
		\State \textit{simultEpiSet} $\leftarrow$ \textit{simultEpiSet} $\cup$  \{$e$ $\mid$ simultaneous event $e$ occurs at $T_e$ and $e$ is after the last event in $e$\};
		\EndFor

		\For{each 1-event/episode  \textit{e} $\in$ \textit{simultEpiSet}}
		\State  simultaneous episode $\beta$ $\leftarrow$ \textbf{\textit{Simult-Concatenate($\alpha$, $e$})}, and get its \textit{moSet}($\beta$);
		
		\State calculate the overall utility and \textit{EWU} of $\beta$ based on $S^\prime$ and \textit{moSet}($\beta$);

		\If {\textit{EWU}($\beta$) $\geq$ the \textit{minUtil}$_{current}$}
		
		\State call \textbf{RUC}\textbf{($\beta$, $k$)};
		
		\State call \textbf{Span-SimultHUE}\textbf{($\beta$, $S^\prime$, \textit{MTD}, \textit{k})};
		
		\State call \textbf{Span-SerialHUE}\textbf{($\beta$, $S^\prime$, \textit{MTD}, \textit{k})};
		\EndIf
		
		\EndFor 		
		
		\State \textbf{return} \textit{top-k HUEs}
	\end{algorithmic}	
\end{algorithm}

In Line 8, we utilize the \textit{EWU}$_{opt}$ pruning strategy to determine whether episode $\beta$ should be retained. If the overall utility of $\beta$ is greater than or equal to the current minimal utility value, this episode will be seen as a new prefix and explored for the next extension (Lines 16--17). All of the above steps aim to filter unpromising episodes and reduce the cost of memory and running time because of the huge candidate searching space. Finally, it returns the set of top-$k$ HUEs that have the common prefix $\alpha$ (Line 20). However, the \textit{Span-SerialHUE} procedure has different operations. The $S$-Concatenate selection of  extension events to construct super-episodes is not the same as in $I$-Concatenate. Due to the intrinsic sequence order and complexity, the number of combinations of episodes is quite large in $S$-Concatenate. In \textbf{Algorithm \ref{Serial-search}}, Lines 2--6 show that from the $T_{\rm e}$+1 timestamp of $\alpha$ to $T_{\rm s}$ + \textit{MTD} time point, all of these events/episodes are candidates for extending the current episode $\alpha$. After expanding an event, the length of $\alpha$ is increased by 1. Further, by recursively using the depth-first search mechanism, candidates will become large. This is a difficult task, but our \textit{EWU}$_{opt}$ can resolve this problem, and this is the reason that we have introduced it in the previous content. After the extension work, Lines 7--21 are the same as those in the \textit{Span-SimultHUE} procedure. Finally, both $I$-Concatenation and $S$-Concatenation share the prefix $\alpha$, and they are only allowed to differ in their last event or element. We also find an interesting phenomenon: if we process $S$-Concatenation first, then process $I$-concatenation, the memory cost is significantly decreased. Obviously, the quantity of candidate generation is significantly increased in corresponding, but the same duration of running time is spent. This is discussed in the next section. In fact, both $I$-Concatenation and $S$-Concatenation share the prefix $\alpha$, and they are only allowed to differ in their last event or element. 

\renewcommand{\algorithmicrequire}{\textbf{Input:}}
\renewcommand{\algorithmicensure}{\textbf{Output:}}
\begin{algorithm}
	\caption{The Span-SerialHUE procedure}
	\label{Serial-search}
	
	\begin{algorithmic}[1]	
		\Require $\alpha$: the current episode; $S^\prime$: the transformed event sequence; \textit{MTD}: user-specified maximum time duration; \textit{k}: desired number of episodes.
		\Ensure \textit{top-k HUEs}: a set of top-$k$ high-utility episodes having $\alpha$ as prefix.
		
		\State initialize \textit{serialEpiSet} = $\phi$;
		
		\For{each \textit{mo}($\alpha$) = $[T_s, T_e] \in$ \textit{moSet}($\alpha$)}
		
		\For{each time point $t$ in [$T_{e}$ + 1, $T_{s}$ + \textit{MTD}]}
		\State \textit{serialEpiSet} $\leftarrow$ \textit{serialEpiSet} $\cup$ \{$e$ $|$ serial event $e$ occurs at $t$\};
		\EndFor
		
		\EndFor
		
		\For{each 1-event/episode $e$ $\in$ \textit{serialEpiSet}}
		
		\State get serial episode $\beta$ $\leftarrow$ \textbf{\textit{Serial-Concatenate($\alpha$, $e$})}, and calculate its \textit{moSet}($\beta$);
		
		\State compute the total utility and \textit{EWU} of $\beta$ based on $S^\prime$ and \textit{moSet}($\beta$);

		\If { \textit{EWU}($\beta$) $\geq$ the \textit{minUtil}$_{current}$}	
		
		\State call \textbf{RUC}\textbf{($\beta$, $k$)};
		
		\State call \textbf{Span-SimultHUE}\textbf{($\beta$, $S^\prime$, \textit{MTD}, \textit{k})};		
		\State call \textbf{Span-SerialHUE}\textbf{($\beta$, $S^\prime$, \textit{MTD}, \textit{k})};
		
		\EndIf
		
		\EndFor 		
		
		\State \textbf{return} \textit{top-k HUEs}
	\end{algorithmic}	
\end{algorithm}

\section{Experimental Study}
\label{sec:experiment}

We conducted experiments on several real-world and synthetic datasets to evaluate the performance of our proposed THUE algorithm. All experiments were performed on a workstation with a 3.0-GHz Intel Core processor and 16 GB of memory, running on Windows 10 Home Edition (64-bit operating system). For the effectiveness evaluation, we compared THUE with the state-of-the-art algorithm TUP \cite{rathore2016top}. We implemented the algorithms in Java using JDK 14; and Rathore \textit{et al.} \cite{rathore2016top} have provided the source code of TUP. The details of the extensive experiments are given in the following sections.

For a more comprehensive analysis of THUE performance, we designed three versions of the designed THUE algorithm in our experiments: 1) THUE$_{ewu}$ denotes the proposed algorithm without the RIU raising threshold strategy (see Strategy 3 in Section 4.3); 2) THUE$_{rus}$ denotes adopting the RUS strategy, but does not utilize the optimized \textit{EWU} pruning strategy; and 3) THUE is the complete algorithm that utilizes all strategies.

\subsection{Datasets and Data Preprocessing}

First, we verified THUE on four datasets, including one synthetic dataset \textit{T10I4D100K}\footnote{\url{http://www.philippe-fournier-viger.com/spmf/index.php?link=datasets.php}}, and three real-world datasets (\textit{Retail}\footnote{\url{http://fimi.ua.ac.be/data/}}, \textit{Foodmart}\footnote{\url{http://msdn.microsoft.com/enus/library/aa217032(v=sql.80).asp}} and \textit{Chainstore}\footnote{\url{http://cucis.ece.northwestern.edu/projects/DMS/MineBench.html}}). These datasets represent the main categories of data with varied features in real-world scenarios. It should be noted that all these datasets were transactional databases. However, they can also be regarded as a long complex event sequence by considering each item as an event, and each transaction is a simultaneous event set at any timestamp. It should be noted that the larger dataset will require more time. What's worse, our equipment has small memory, so the result value will be huge. For all datasets except Foodmart, we took the first 10,000 transactions as experimental testing. If not explicitly specified, we always fix the maximum time duration (\textit{MTD}) parameter to 2 for our experiments. The characteristics of the datasets above are described in Table \ref{table:datasets}.

\begin{table}[htb]
	\centering
	\caption{Characteristics of datasets}
	\label{table:datasets}
	\begin{tabular}{lcccc} \hline
		\textbf{Dataset}  & \textbf{\#Trans} & \textbf{Avg.length} & \textbf{\#Items} & \textbf{Type} \\
		\hline
		Retail		& 10,000 & 5.2 & 16,470 & Sparse \\
		Foodmart	& 4,141  & 4.4 & 1,559  & Sparse \\
		T10I4D100K  & 10,000 & 10  & 870    & Sparse \\
		Chainstore  & 10,000 & 7   & 46,085 & Sparse \\
		\hline
	\end{tabular}
\end{table}

More details of these datasets can be found in the SPMF \cite{fournier2014spmf} open-source library. All the datasets have been published and are available to researchers. Similar to previous studies \cite{rathore2016top,gan2019discovering}, we use a simulation method to randomly generate the internal and external utilities in datasets, except Foodmart and T10I4D100K, as follows: 1) generate the internal utility (in the range of 1 to 5) for each item in every transaction; 2) set the external utility for each item (in the range of 1 to 1,000 using a log-normal distribution). The original T10I4D100K dataset contained negative items. Therefore, we only take absolute values as a new dataset and do not change its structure. We test each method and show the final results (i.e., runtime and memory consumption). When the runtime exceeds 10,000 s or out of memory, we suppose that there is no result of the experimental algorithm. Thus, the results of related patterns are marked as ``-" in tables and ``0" in figures. In particular, it should be noted that TUP cannot obtain results within 10,000 s in most datasets because of the huge episodes generated and its inefficiency. Thus, most of its test results are marked as ``-,¡± and we mainly analyze the performance of the other three algorithms.

\subsection{Effectiveness Evaluation}

\begin{figure*}[htbp]
	\centering 
	\includegraphics[trim=60 0 15 10, clip,height=0.17\textheight, width=1.02\textwidth]{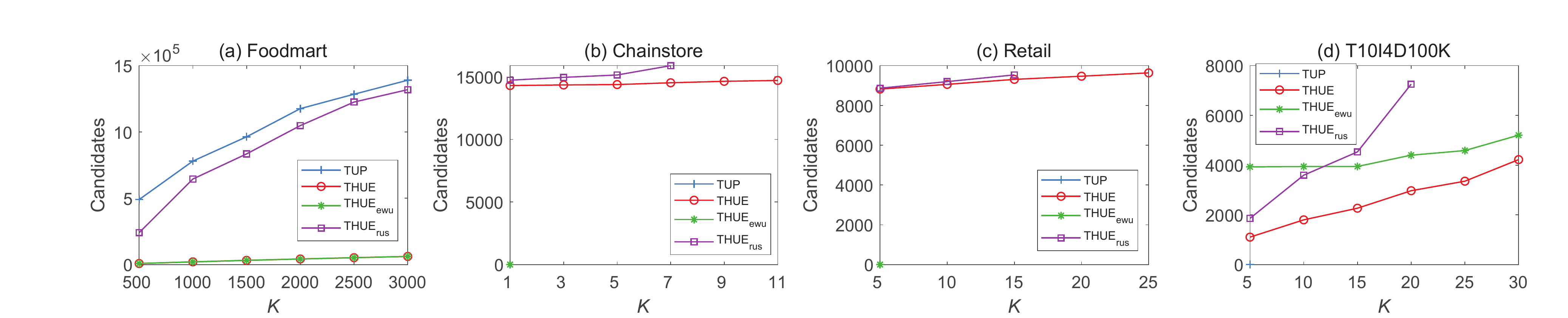}
	\caption{Candidates generated under various parameters (\textit{MTD} and \textit{minUtil}).}
	\label{fig:Candidates}	
\end{figure*}

As shown in Fig. \ref{fig:Candidates}, the tendency of the candidate generation is very clear. We use the running example as a case study to evaluate the generation candidates during the mining process discovered by the state-of-the-art TUP \cite{rathore2016top}, THUE$_{ewu}$, THUE$_{rus}$, and THUE algorithms. The mining results from the event sequence in four (Retail, T10I4D100K, Chainstore, and Foodmart) datasets are plotted in Table \ref{table:candidates}. It is first observed that as $k$ increases, the number of candidates generated also increases. This is reasonable because as more episodes are found, more episodes are considered as potential HUEs before \textit{minUtil} updates. The number of candidates generated represents the efficiency of the algorithm. According to obverse candidate information, we can initially deduce which one requires less runtime and memory consumption than the other comparison algorithms. Furthermore, the worst algorithm can also be easily identified. For instance, in Table \ref{table:candidates}, it is easy to observe that TUP produces 490,214 candidates, while the others produce 7,947, 240,288, and 7,887, respectively. It always runs overtime or is out of memory in the other experiments. Thus, we suppose that TUP is the worst algorithm, and our proposed novel algorithm works better than TUP.

Second, both TUP and THUE$_{ewu}$ fail to quickly extract the complete true high-utility episodes from the event sequence. For example, the average quantity of candidate TUP products is more than 30 times that of THUE in Foodmart. This is because TUP computes the original \textit{EWU} upper bound and does not consider removing the overlapping utilities at $T_{\rm e}$. In contrast, the pruning strategy \ref{strategy:EWU} based on the optimized \textit{EWU} that THUE adopts cuts many low-utility episodes to avoid extending those nodes. At the same time, THUE$_{ewu}$ produces almost the same number as TUP in the other datasets. For instance, in the Retail and Chainstore datasets, both run overtime, even when $k$ is the smallest. The reason for this is that the user-specified variable $k$ is too small, and the magnitude of the $k$ values does not change much after the mining process. Furthermore, THUE and THUE$_{rus}$ both adopt Strategies \ref{strategy:riu} and \ref{strategy:rtu}, the threshold is raised significantly before searching the LS-tree, and the difference between the number of candidates generated is not very large (less than 1,000). Although THUE$_{rus}$ also utilizes efficient raising strategies, it still runs overtime in T10I4D100K ($k$ = 25, 30) and Chainstore ($k$ = 9, 11). An explanation for this is provided in the next subsection.

\begin{table}[htbp]
	\fontsize{5pt}{9pt}\selectfont
	\centering
	\caption{Generated candidates by four top-$k$ HUEM algorithms}
	\label{table:candidates}
	\begin{tabular}{|c|c|llllll|}
		\hline\hline
		\textbf{Dataset} & \textbf{Algorithm} & \multicolumn{6}{c|}{\textbf{\# Candidates when varying $k$ with \textit{MTD} = 2}}\\ \hline
		
		&\textit{\textbf{k}}	   & 5 & 10 & 15 & 20 & 25 & 30 \\ \cline{2-8}
		&\textbf{TUP} 		   & - & - & - & - & - & - \\
		\textbf{Retail}&\textbf{THUE$_{ewu}$}  & - & - & - & - & - & - \\
		&\textbf{THUE$_{rus}$}  & 8,866 & 9,198 & 9,535 &   -   &   -   & - \\
		&\textbf{THUE}   	   & 8,822 & 9,058 & 9,310 & 9,469 & 9,636 & - \\ \hline
		
		&\textit{\textbf{k}}	   & 5 & 10 & 15 & 20 & 25 & 30 \\ \cline{2-8}
		&\textbf{TUP} 		   & - & - & - & - & - & - \\
		\textbf{T10I4D100K}&\textbf{THUE$_{ewu}$}  & 3,930 & 3,942 & 3,925 & 4,399 & 4,585 & 5,203 \\
		&\textbf{THUE$_{rus}$}  & 1,858 & 3,594 & 4,531 & 7,255 &   -   &   -   \\
		&\textbf{THUE}   	   & 1,103 & 1,798 & 2,266 & 2,970 & 3,353 & 4,221 \\ \hline
		
		&\textit{\textbf{k}}	   & 1 & 3 & 5 & 7 & 9 & 11 \\ \cline{2-8}						
		&\textbf{TUP} 		   & - & - & - & - & - & - \\
		\textbf{Chainstore}&\textbf{THUE$_{ewu}$}  & - & - & - & - & - & - \\
		&\textbf{THUE$_{rus}$}  & 14,722 & 14,949 & 15,137 & 15,884 & 	-   & 	- 	 \\
		&\textbf{THUE}   	   & 14,293 & 14,342 & 14,374 & 14,516 & 14,633 & 14,702 \\ \hline
		
		&\textit{\textbf{k}}	 & 500 & 1000 & 1500 & 2000 & 2500 & 3000 \\ \cline{2-8}	
		&\textbf{TUP} 			 & 490,214 & 780,297 & 963,478 & 1,175,965 & 1,284,491 & 1,390,133 \\
		\textbf{Foodmart}&\textbf{THUE$_{ewu}$}  & 7,947   & 20,253  & 31,569  & 42,007    & 51,806    & 61,212    \\
		&\textbf{THUE$_{rus}$}  & 240,288 & 644,969 & 834,477 & 1,047,827 & 1,225,352 & 1,318,996 \\
		&\textbf{THUE}   		 & 7,887   & 20,241  & 31,569  & 42,007    & 51,806    & 61,212    \\ \hline
		
		\hline\hline
	\end{tabular}
\end{table}

\subsection{Runtime Analysis}

\begin{figure*}[htbp]
	\centering 
	\includegraphics[clip,height=0.15\textheight, width=1.02\textwidth]{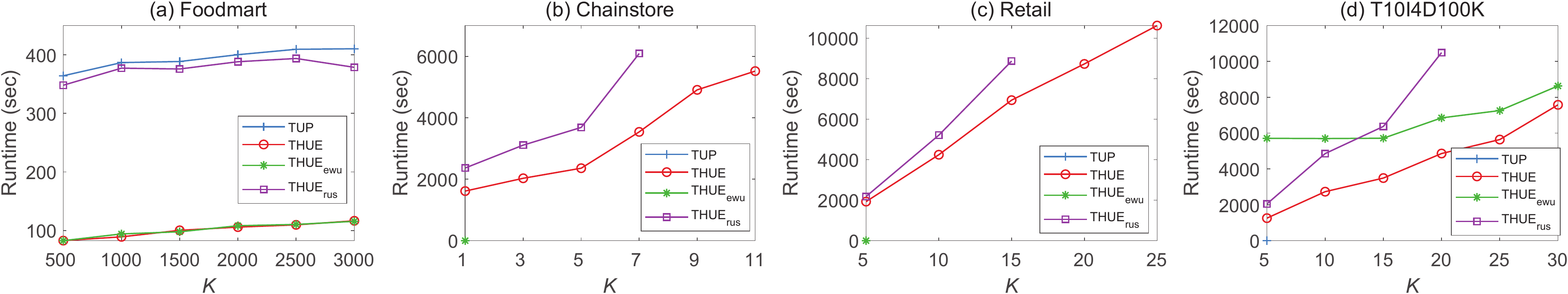}
	\caption{Runtime under various parameters ($K$).}
	\label{fig:Runtime}	
\end{figure*}

A runtime analysis is introduced in this subsection. ``Runtime" indicates the tested running time of each variant of the THUE algorithm, by varying $k$. Note that the scale of runtime is in seconds. In Fig. \ref{fig:Runtime}, it can be observed that the performance of TUP is the worst among all the algorithms. This always runs overtime in our experiments. This is because TUP does not utilize efficient increasing threshold strategies before constructing $l$-episodes (l $\ge$ 2); it also spends too much time searching HUEs. In addition, the complexity of concatenation operations in sequence data-mining algorithms has become more difficult.

We can also learn from each sub-figure that the total execution time of each compared algorithm is highly related to the values of $k$. Although we set $k$ to be very small (e.g., 1, 3, 5, etc.), THUE still requires thousands of seconds of runtime, except for the results shown in Fig. \ref{fig:Runtime}(a). In addition, the differences in the runtime of the three comparison algorithms were mainly related to the amount of generated promising episodes. For example, in Fig. \ref{fig:Runtime}(c), the runtime increases from approximately 2,000 s to 10,000 s, while $k$ grows from 5 to 25. When $k$ is 25, THUE requires nearly 10,000 s; then, we suppose that THUE must be overtime when $k$ = 30. Therefore, it is easy to understand why THUE$_{ewu}$ and THUE$_{rus}$ can only obtain a few results at the same time. In the case of T10I4D100K, as shown in Fig. \ref{fig:Runtime}(d), we can observe the difference of the execution time between THUE and THUE$_{rus}$ and its trend. When $k$ is set to 20 on the T10I4D100K dataset, THUE$_{rus}$ requires approximately twice as much time as THUE.

Fig. \ref{fig:Runtime}(a) shows that as the $k$ values become larger, the advantage of the optimized \textit{EWU} pruning strategy is gradually revealed. THUE$_{rus}$, which does not adopt the optimized \textit{EWU} technique, requires three times the cost of THUE and THUE$_{ewu}$. As shown in Fig. \ref{fig:Candidates}(a), THUE$_{rus}$ often generates over one million candidates, while THUE produces 10,000. We should point out that especially for the dataset where the average transaction length is large, a high $k$ can dramatically increase the running time. In addition, small changes around the \textit{MTD} may have a noticeable effect on the runtime, as shown in Fig. \ref{fig:Scalability}.

\subsection{Memory Cost Analysis}

\begin{figure*}[htbp]
	\centering 
	\includegraphics[clip,height=0.15\textheight, width=1.0\textwidth]{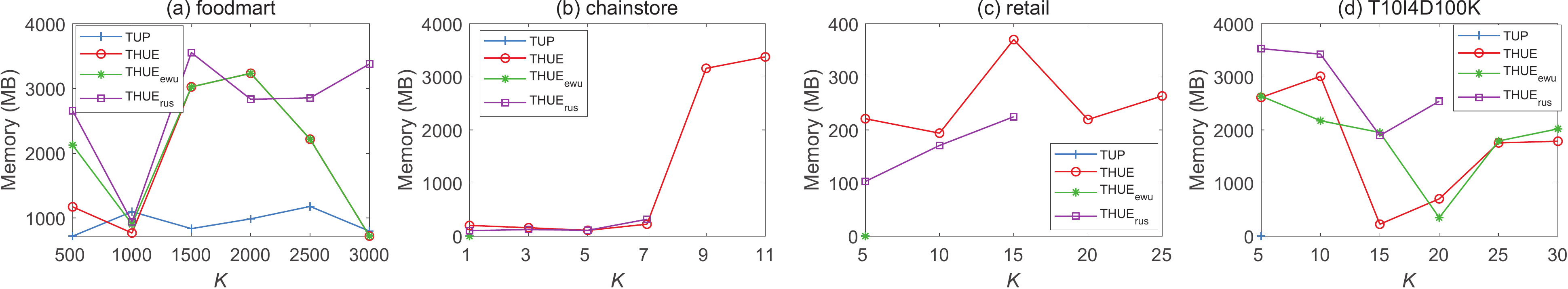}
	\caption{Memory cost under various values of $K$.}
	\label{fig:Memory}	
\end{figure*}

In this series of experiments, we evaluate the memory consumption performance of our proposed algorithms with different $k$ values for mining HUEs. Fig. \ref{fig:Memory} shows the memory consumption under varying the $k$ with the fixed size of the target datasets. We then analyze the memory cost performance of our proposed algorithms. It should be pointed out that the missing values represent runtime exceeding 10,000 s and learning the corresponding records, as shown in Fig. \ref{fig:Runtime}.

In the Chainstore dataset, TUP, THUE$_{ewu}$, and THUE$_{rus}$ always run overtime because we can see that the memory consumption of THUE increases rapidly after $k$ = 7. The reason for this is a feature of the dataset. In the Foodmart dataset, it seems that TUP consumes much less memory than the other algorithms under varying $k$. However, in Fig. \ref{fig:Runtime}(a), we can observe that TUP has the highest runtime. Table \ref{table:threshold} shows that the final thresholds are always less than those of the other algorithms. Except for the Foodmart dataset, the results of the memory consumption of the algorithms are directly related to the number of candidates they generate. Table \ref{table:threshold} provides the details.

Note that, except for Chainstore, the memory consumption of THUE does not seem to be very good. For example, THUE always requires more memory than THUE$_{rus}$ under varying $k$ in retail. Furthermore, THUE is the only one that can obtain all results with different $k$ parameters in all datasets. The others often run overtime with no output. As shown in Fig. \ref{fig:Candidates} and Fig. \ref{fig:Runtime}, THUE can speed up processing when spanning the LS-tree and reduce intermediate candidate patterns significantly. In our candidate experiments, the THUE algorithm performed best in all datasets.

\subsection{Threshold Comparison}

\begin{figure}[htbp]
	\centering 
	\includegraphics[clip,width=0.5\textwidth]{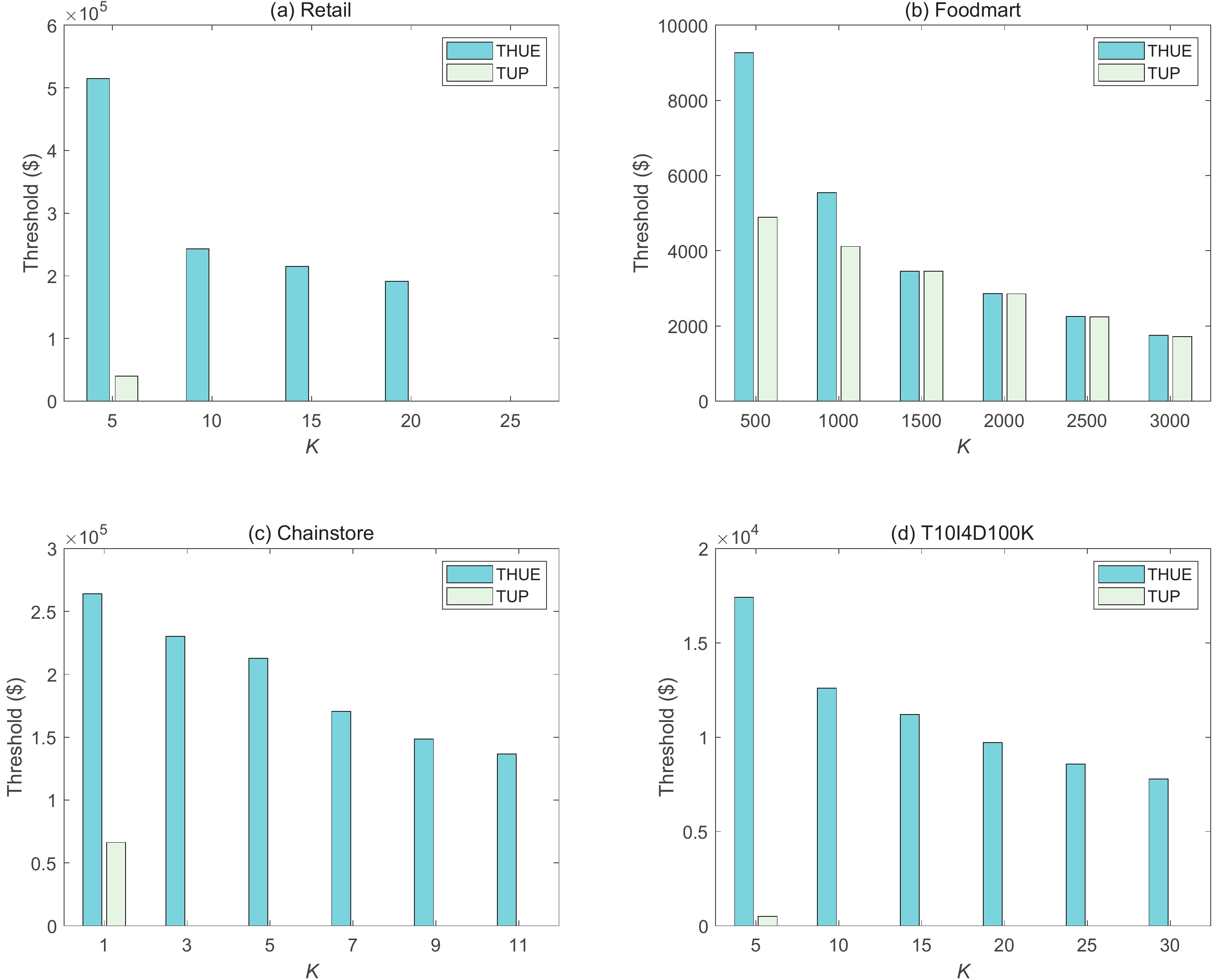}
	\caption{Initial \textit{minUtil} for various values of $K$.}
	\label{fig:MinUtil}	
\end{figure}

FIg. \ref{fig:MinUtil} shows the initial \textit{minUtil} comparison between TUP and THUE. And Table \ref{table:threshold} lists the detail changes of \textit{minUtil} after the Foodmart and T10I4D100K datasets experiment. ``Initial" represents \textit{minUtil}, where the algorithms first use the increasing threshold strategy after scanning the dataset. ``End" means the last \textit{minUtil} value when experiment is terminated. We discover that more HUEs, initial and final \textit{minUtil}, are continuously decreasing because the value of $k$ increases. Compared with the initial and final \textit{minUtil}, THUE increases up to one order of magnitude. We find that the final \textit{minUtil} of TUP is slightly less than that of the other algorithms in the first row (Foodmart dataset). The reason may be a bug in the TUP or other uncontrollable factors, and this scenario does not affect our analysis of the results. THUE$_{ewu}$ utilizes the same raising threshold methods of TUP, and thus the initial \textit{minUtil} values are always equal.

\begin{table}[htb]
	\fontsize{5pt}{9pt}\selectfont
	\centering
	\caption{Threshold comparison of four top-$k$ HUEM algorithms}
	\label{table:threshold}
	\begin{tabular}{|c|c|cllllll|}
		\hline\hline
		\textbf{Dataset} & \textbf{Algorithm} & \multicolumn{7}{c|}{\textbf{\# minUtil (\$) when varying $k$ with \textit{MTD} = 2}}\\ \hline
		
		&\textit{\textbf{k}}	   	   &  		 & 500 	 & 1000  & 1500  & 2000  & 2500  & 3000   \\ \cline{2-9}
		&\multirow{2}{*}{\textbf{TUP}}& initial & 4,887 & 4,114 & 3,453 & 2,854 & 2,244 & 1,715  \\
		& 							   & end 	 & 12,284& 12,023& 11,860& 11,741& 11,647& 11,573 \\ \cline{2-9}
		&\multirow{2}{*}{\textbf{THUE$_{ewu}$}}& initial & 4,887 & 4,114 & 3,453 & 2,854 & 2,244 & 1,715  \\
		\textbf{Foodmart}& 							   		    & end	  & 12,516& 12,136& 11,937& 11,802& 11,698& 11,614 \\ \cline{2-9}
		&\multirow{2}{*}{\textbf{THUE$_{rus}$}}& initial & 9,272 & 5,544 & 3,456 & 2,856 & 2,257 & 1,752  \\
		& 							   		    & end	  & 12,516& 12,136& 11,937& 11,802& 11,698& 11,614 \\ \cline{2-9}
		&\multirow{2}{*}{\textbf{THUE}}& initial & 9,272 & 5,544 & 3,456 & 2,856 & 2,257 & 1,752  \\
		& 							    & end	  & 12,516& 12,136& 11,937& 11,802& 11,698& 11,614 \\ \hline
		
		
		&\textit{\textbf{k}}	   		 &  	   & 5 & 10 & 15 & 20 & 25 & 30 \\ \cline{2-9}
		&\multirow{2}{*}{\textbf{TUP}}& Initial &494& -  & -  & -  & -  & -  \\
		& 							 & end 	   & - & -  & -  & -  & -  & -  \\ \cline{2-9}
		&\multirow{2}{*}{\textbf{THUE$_{ewu}$}}& initial & 494   & 468   & 456 	 & 448 	  & 439   & 436  \\
		\textbf{T10I4D100K}& 							   		  & end		& 11,418& 11,418& 11,418 & 10,030 & 9,660 & 8808 \\ \cline{2-9}
		&\multirow{2}{*}{\textbf{THUE$_{rus}$}}& initial & 17,410 & 12,600 & 11,200 & 9,716  & - & - \\
		& 							   		  & end		& 17,410 & 12,600 & 11,418 & 10,030 & - & - \\ \cline{2-9}
		&\multirow{2}{*}{\textbf{THUE}}& initial & 17,410 & 12,600 & 11,200 & 9,716 & 8,572 & 7,785 \\
		& 							  & end		& 17,410 & 12,600 & 11,418 & 10,030& 9,660 & 8,808 \\ \hline
		
		\hline\hline
	\end{tabular}
\end{table}

In the second row of Table \ref{table:threshold} (T10I4D100K dataset), when $k$ is set to 5, the gap between the initial \textit{minUtils} of TUP and THUE is so large that TUP runs overtime, and later we do not need to continue to test TUP. There is another interesting scene where THUE$_{ewu}$ has the same final \textit{minUtil} when $k$ is 5 and 10, respectively. We believe that THUE$_{ewu}$ only adopts the RTU and RUC strategies, which cannot increase the threshold rapidly, and this will make some final \textit{minUtil} values inaccurate. Because the value of $k$ is small, we can also observe that \textit{minUtil} of THUE does not change significantly. In fact, most HUEs are 1-episode because of the large initial \textit{minUtil}.

\subsection{Scalability Test}

\begin{figure}[htbp]
	\centering  
	\includegraphics[clip,width=0.5\textwidth]{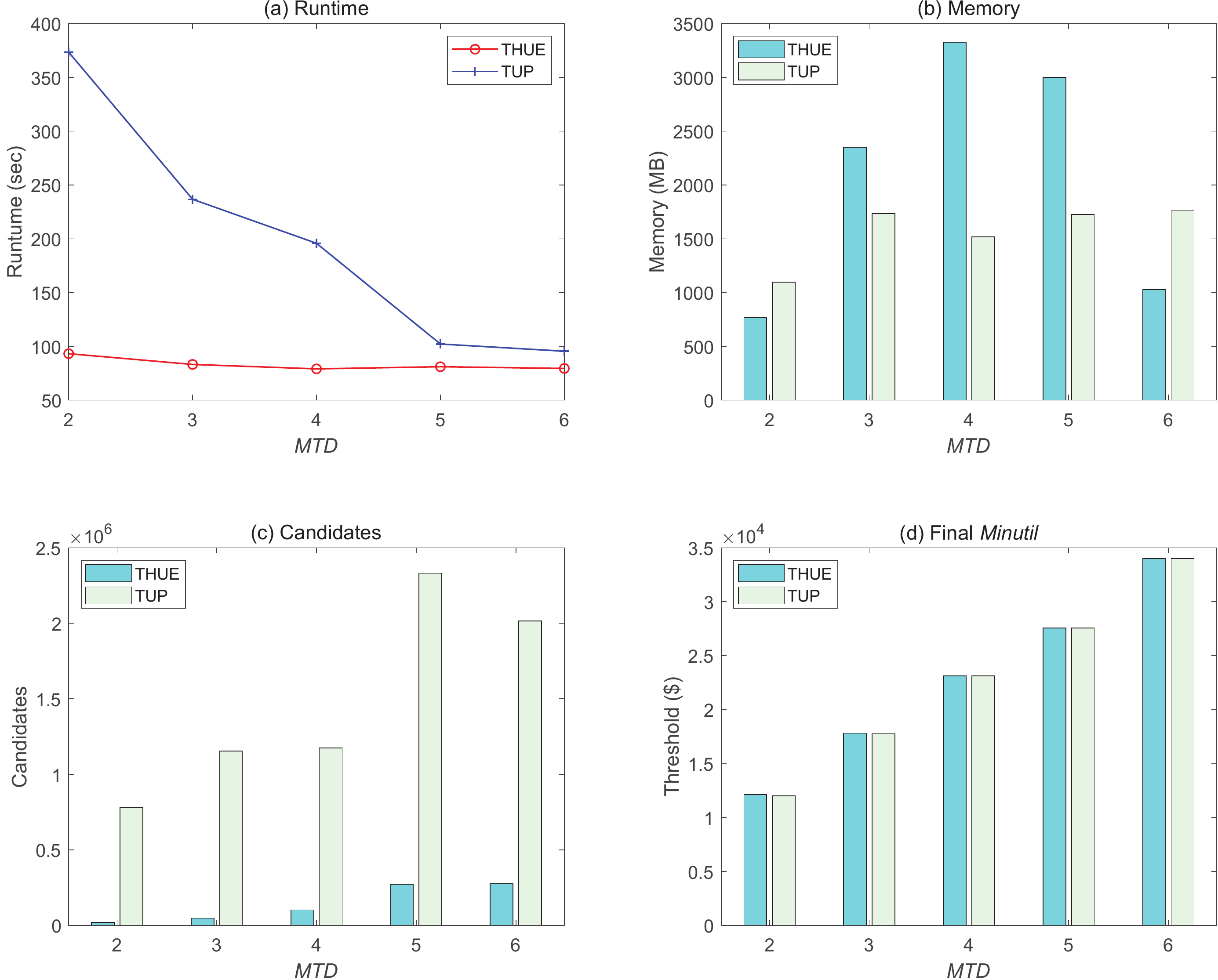}
	\caption{Scalability test with different \textit{MTD}.}
	\label{fig:Scalability}	
\end{figure}

The most significant problem of HUEM is that it is more computationally efficient because of the long event sequence. Fig. \ref{fig:Scalability} shows the results regarding the scalability of comparison between TUP and THUE with different \textit{MTDs} (from 2 to 6) in a complex event sequence. We selected Foodmart as the test dataset, and parameter $k$ was set to 1,000.  It is shown that the execution time changes slightly with respect to varying \textit{MTDs}. As the memory cost increases, the number of intermediate candidates increases very slowly. Compared to the performance of TUP, the memory consumption is less than THUE, whereas the quantity of intermediate candidates and runtime cost has a huge gap in THUE. Also note that the execution time of TUP decreases gradually. With a fixed $k$ and dataset, we refer to the larger \textit{MTD} value and the smaller episodes \textit{minUtil}. We can clearly see that the final \textit{minUtil} gradually increases, and thus may prune more low-utility episodes more easily.

\section{Conclusion}
\label{sec:conclusions}

In this study, we incorporated the concept of top-$k$ high-utility episode mining. Because of the complexity of episode concatenation, candidate generation is a challenge that cannot be ignored. The novel algorithm THUE was updated using the UMEpi algorithm. We utilized some automatic minimal utility raising strategies (RUS and RUC), and combined them with the powerful pruning strategy, which is based on optimized episode-weighted utilization (\textit{EWU}). Extensive experiments on some synthetic and real-world datasets demonstrated that THUE improves episode mining efficiently and cuts many unnecessary operations. According to the top $k$ idea, users can avoid wasting a lot of time to find a proper minimum utility threshold.

In future work, we would like to discover other types of episodes, such as parallel and closed episodes. Furthermore, we plan to design several more useful \textit{minUtil} strategies to obtain better performance. Finally, the design of a distributed model \cite{gan2017data} of THUE is also interesting and challenging.

\ifCLASSOPTIONcaptionsoff
  \newpage
\fi

\bibliographystyle{IEEEtran}
\bibliography{THUE}


\vspace{-1.5cm}
\begin{IEEEbiography}[{\includegraphics[width=1in,height=1.25in,clip,keepaspectratio]{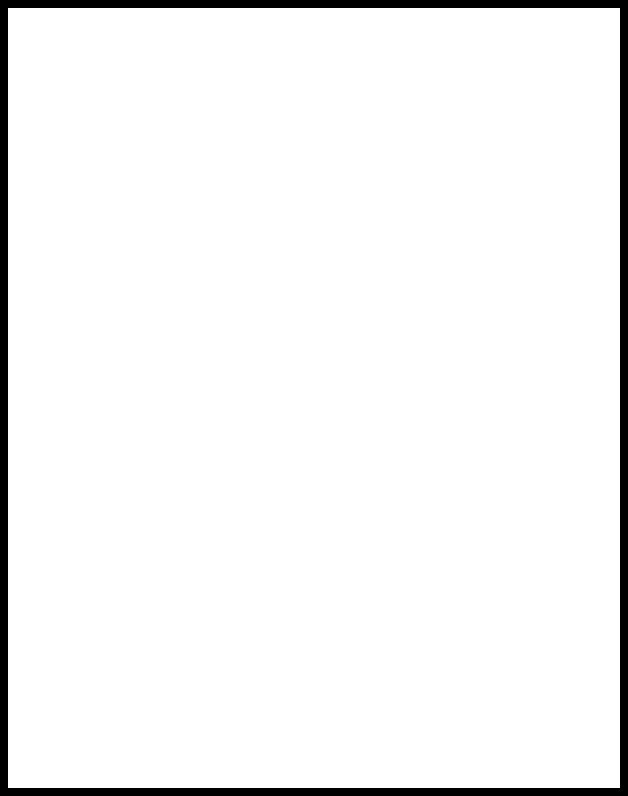}}]{Shicheng Wan} 
	received the B.S. degree in Gannan Normal University, Ganzhou, China in 2020. He is currently a master student with the Department of Computer Sciences, Guangdong Technology University, China. His research interests include data mining, utility mining, and big data. 
\end{IEEEbiography}

\vspace{-1.5cm}
\begin{IEEEbiography}[{\includegraphics[width=1in,height=1.25in,clip,keepaspectratio]{newAuthor.png}}]{Jiahui Chen (Member, IEEE)}
	received the BS degree from South China Normal University, China, in 2009, and MS and PhD degrees from South China University of Technology, China, in 2012 and 2016, respectively. He joined National University of Singapore as a research scientist between form 2017 to 2018. He is currently an associate professor in the School of Computer at Guangdong University of Technology. His research interests mainly focus on public key cryptography, post-quantum cryptography, and information security. 
\end{IEEEbiography}

\vspace{-1.5cm}
\begin{IEEEbiography}[{\includegraphics[width=1in,height=1.25in,clip,keepaspectratio]{newAuthor.png}}]{Wensheng Gan  (Member, IEEE)} 
	received the B.S. degree in Computer Science from South China Normal University, China in 2013. He received the Ph.D. in Computer Science and Technology, Harbin Institute of Technology (Shenzhen), China in 2019. He was a joint Ph.D. student with the University of Illinois at Chicago, Chicago, USA, from 2017 to 2019. He is currently an Association Professor with the College of Cyber Security, Jinan University, Guangzhou, China.  His research interests include data mining, utility computing, and big data analytics. He has published more than 80 research papers in peer-reviewed journals (i.e., IEEE TKDE, IEEE TCYB, ACM TKDD, ACM TOIT, ACM TMIS) and international conferences. He is an Associate Editor of \textit{Journal of Internet Technology}. 
\end{IEEEbiography}

\vspace{-1.5cm}
\begin{IEEEbiography}[{\includegraphics[width=1in,height=1.25in,clip,keepaspectratio]{newAuthor.png}}]{Guoting Chen}  
	  received B.S., M.S. and Ph.D. degrees in Mathematics from Wuhan University, China in 1982, from Wuhan University, China in 1985, and from University de Grenoble 1, France in 1990, respectively. He is currently a full professor with School of Science, Harbin Institute of Technology, Shenzhen.  His research interests include Mathematics, differential equations, and data science. He has published 30 peer-reviewed research papers. 
\end{IEEEbiography}

\vspace{-1.5cm}
\begin{IEEEbiography}[{\includegraphics[width=1in,height=1.25in,clip,keepaspectratio]{newAuthor.png}}]{Vikram Goyal}  
	received the PhD in Computer Science and Engineering from the Department of Computer Science and Engineering at IIT Delhi in 2009. Before pursuing PhD, he completed MTech in Information Systems from the Department of Computer Science and Engineering at NSIT Delhi in 2003. He has published 50  research papers in conferences and referred journals. He has a couple of Projects from DST, India and Deity, India on problems related to Privacy in Location-based Services and Digitized Document Fraud Detection, respectively.
\end{IEEEbiography}




\end{document}